\documentclass[fleqn,usenatbib]{mnras}

\usepackage{newtxtext,newtxmath}
\usepackage[T1]{fontenc}

\DeclareRobustCommand{\VAN}[3]{#2}
\let\VANthebibliography\thebibliography
\def\thebibliography{\DeclareRobustCommand{\VAN}[3]{##3}\VANthebibliography}

\usepackage[pagewise]{lineno}

\usepackage[caption=false]{subfig}
\usepackage{graphicx}	% Including figure files
\usepackage{amsmath}	% Advanced maths commands

\usepackage{comment}

\title[MeerKLASS L-band DR1]{The MeerKLASS L-band On-the-Fly Continuum Survey: Data Release 1}

\author[S. Mangla et al.]{Sarvesh Mangla$^{1}$\thanks{E-mail: mangla.sarvesh@physik.lmu.de},
Joseph J. Mohr,$^{1}$,
Kristof Rozgonyi$^{1}$,
Suman Chatterjee$^2$,
Keith Grainge$^{3}$,
\newauthor
Sourabh Paul$^3$,
Mario G. Santos$^{2,4}$,
Yvette Perrott$^{5}$,
Oleg M. Smirnov$^{4,6,7}$,
Cyril Tasse$^{8,9}$,
Laura Wolz$^3$
\\
\\
$^{1}$University Observatory, Faculty of Physics,  Ludwig-Maximilians-Universität, Scheinerstr. 1, 81679, Munich, Germany\\
$^2$Department of Physics and Astronomy, University of the Western Cape, Robert Sobukwe Road, Cape Town 7535, South Africa\\
$^3$Jodrell Bank Centre for Astrophysics, Department of Physics \& Astronomy, The University of Manchester, Manchester M13 9PL, UK\\
$^4$South African Radio Astronomy Observatory (SARAO), 2 Fir Street, Cape Town, 7925, South Africa\\
$^{5}$School of Chemical and Physical Sciences, Victoria University of Wellington, Wellington 6012, New Zealand\\
$^6$Centre for Radio Astronomy Techniques and Technologies (RATT), Department of Physics and Electronics, Rhodes University, Makhanda, 6140, South Africa\\
$^7$Institute for Radioastronomy, National Institute of Astrophysics (INAF IRA), Via Gobetti 101, 40129 Bologna, Italy\\
$^8$GEPI \& ORN, Observatoire de Paris, Université PSL, CNRS, 5 Place Jules Janssen, 92190 Meudon, France\\
$^9$Department of Physics \& Electronics, Rhodes University, PO Box 94, Grahamstown, 6140, South Africa\\
}

\date{Accepted XXX. Received YYY; in original form ZZZ}

\pubyear{\the\year{}}

\begin{document}
\label{firstpage}
\pagerange{\pageref{firstpage}--\pageref{lastpage}}
\maketitle

% Abstract of the paper
\begin{abstract}

The MeerKAT Large Area Synoptic Survey (MeerKLASS) collaboration has acquired multiple passes of L-band (856-1712\,MHz) scanning observations over a 268\,deg$^2$ sky region.  This scanning enables efficient, large-area sky surveys by continuously scanning the MeerKAT array back and forth at fixed elevation while recording data at 2\,s intervals, progressively covering the survey region as the Earth rotates.
We employ a novel on-the-fly (OTF) interferometric imaging technique to construct continuum images and catalogs from 16\,hours of scan data. These data products, constituting the first MeerKLASS L-band data release (DR1), consist of high-fidelity radio continuum images and a catalogue of 34,874 radio sources detected with a SNR$>$9. The resulting Stokes I images achieve a median noise level of $33\,\umu$Jy\,beam$^{-1}$ and a median angular resolution of approximately $25.5\arcsec \times 7.8\arcsec$. Cross-comparisons with previous surveys confirm the consistency of our flux density scale within 5\% and astrometric precision within $1.5\arcsec$. Additionally, flux densities measured across the seven sub-bands enable in-band spectral-index estimates for the detected sources, providing insights into their physical properties and the broader source population. We compute the differential source counts, finding good agreement with existing measurements and validating our end-to-end processing. This data release demonstrates the effectiveness of scanning surveys when combined with OTF interferometric imaging. Commensal intensity mapping and interferometric imaging offers a dramatic enhancement of survey science per invested hour of observations and could therefore be an appealing option for next generation facilities like SKA-Mid.

\end{abstract}

% Select between one and six entries from the list of approved keywords.
% Don't make up new ones.

%%%%%%%%%%%%%%%%%%%%%%%%%%%%%%%%%%%%%%%%%%%%%%%%%%

%%%%%%%%%%%%%%%%% BODY OF PAPER %%%%%%%%%%%%%%%%%%

\section{Introduction}

Conducting ever more sensitive and larger solid angle surveys to explore our Universe has long been a core pursuit in astronomical research. Over the last six decades, radio surveys have steadily progressed in depth, resolution and precision. Yet, the arrival of new, enhanced, and upcoming instruments holds the potential to transform the field dramatically. One such instrument is MeerKAT, which will eventually contribute to the larger SKA-Mid array. The MeerKLASS \citep[MeerKAT Large Area Synoptic Survey;][]{Mario_2016} project is designed to map extensive regions of the sky, primarily for cosmological studies via single-dish HI intensity mapping. One of the key scientific goal of MeerKLASS is to complement this with a commensal large solid angle, high angular-resolution interferometric survey, made possible through the use of `On-the-Fly' (OTF) interferometric imaging. The MeerKLASS commensal scan survey therefore enhances both large-scale structure studies and high-resolution astrophysical investigations of the formation and evolution of massive black holes, galaxies, and clusters of galaxies.

\begin{table*}
\centering
\begin{tabular}{ccccccc}
\hline
Block ID & Rising (R)/Setting (S) & Observation Window (UTC) & Calibrator & OTF Duration \\
\hline %20:08:32 - 18:29:03 
1630519596 & R & 2021-09-01 18:06:52 -- 20:08:23 & 1934$-$638 & 103 min \\ %19:58:36 - 18:15:56
1631387336 & R & 2021-09-11 19:09:12 -- 21:05:50 & 1934$-$638 & 97 min \\ %20:55:53 - 19:18:17
1631552188 & R & 2021-09-13 16:56:48 -- 19:00:08 & 1934$-$638 & 105 min \\ %18:50:20 - 17:05:54
1631724508 & R & 2021-09-15 16:48:48 -- 18:51:59 & 1934$-$638 & 105 min \\ %18:42:13 - 16:57:57
1631982988 & R & 2021-09-18 16:36:46 -- 18:39:05 & 1934$-$638 & 104 min \\ %18:29:22 - 16:45:55
1632077222 & R & 2021-09-19 18:47:17 -- 20:43:56 & 1934$-$638 & 97 min \\ %20:33:58 - 18:56:20
1632760885 & R & 2021-09-27 16:41:42 -- 18:42:56 & 1934$-$638 & 103 min \\ %18:33:05 - 16:50:45
1634835083 & R & 2021-10-21 16:51:42 -- 18:47:17 & 1934$-$638 & 97 min \\ %18:37:01 - 17:00:51
\hline
\end{tabular}
\caption{Summary of the eight MeerKLASS L-band observing blocks included in this data release. Each entry lists the block ID, scan direction, observation start and end times, calibrator source, and effective duration of the OTF science scan.}
\label{tab:obs_summary}
\end{table*}

Radio continuum extragalactic surveys provide a uniquely powerful means to study galaxy formation and evolution across cosmic time.  Unlike observations in the ultraviolet and optical bands, radio wavelengths are largely unaffected by dust obscuration, allowing for an unobstructed view of diverse galaxy populations, especially at high redshifts. Radio signals arises from various astrophysical mechanisms, including synchrotron emission from the compact cores and extended jets of active galactic nuclei (AGN), synchrotron and thermal emission from supernova-driven processes in star-forming galaxies (SFGs) and inverse Compton scattering of the cosmic microwave background (CMB) off populations of energetic electrons in, e.g., clusters of galaxies. While AGN produce strong radio emission, typical SFGs emit fainter synchrotron and thermal radiation, particularly at higher frequencies. Crucially, radio emission from SFGs has been established as a robust indicator of star formation rates \citep[SFR;][]{Yun_2001,Bell_2003,Pannella_2009}, reinforcing the central role of radio continuum studies in understanding the evolution of galaxies.

The MeerKLASS survey is strategically configured to leverage MeerKAT's commensal observing capability. While the primary driver is the single-dish $\text{H}\text{I}$ cosmology, which necessitates a constant-elevation, fast-scanning strategy \citep[e.g.,][]{Mangum_2007}, the MeerKAT array simultaneously records interferometric visibilities enabling OTF imaging. This dual approach unifies the central science objectives: (i) to constrain the growth of cosmic structure and the distance-redshift relation via $\text{H}\text{I}$ intensity mapping, and (ii) to assemble a statistical census of the radio-source population across cosmic time using the deep continuum data. This methodology delivers contiguous, high efficiency and high-fidelity coverage over a large fraction of the extragalactic sky, while enabling a combined approach to cosmological and astrophysical science. The dense core configuration of MeerKAT, combined with its low system temperature and advanced calibration, ensures sensitivity to both diffuse emission and compact sources. 

In single-dish mode, MeerKLASS observes the combined 21\,cm emission from many unresolved galaxies, using the H\,I intensity–mapping technique \citep[e.g.][]{Chang_2010,Masui_2013,Anderson_2018,Wolz_2022,CHIME_2022,Cunnington_2023,Paul_2023}. Rather than detecting individual H\,I galaxies, this approach measures fluctuations in the neutral-hydrogen field on large scales up to $z \lesssim 1.6$. The resulting maps trace the underlying matter distribution and include the baryon acoustic oscillation (BAO) feature, which can be used as a standard ruler to constrain the expansion history of the Universe. In this way, MeerKLASS can provide independent constraints on dark energy and tests of modified gravity, while also serving as a pathfinder for future SKA H\,I cosmology surveys.

Operationally, the constant-elevation scan strategy provides several key advantages for wide-area OTF imaging. It allows for the efficient mapping of extensive sky regions, and by mitigating direction-dependent calibration artefacts often associated with static mosaics, it produces images with uniform thermal noise and a stable synthesized beam (hereafter referred to as the point spread function PSF) across wide fields. This technique, successfully implemented in other interferometric surveys such as the VLA Sky Survey (VLASS; \citealt{Lacy_2020}), has been adapted here to MeerKAT’s L-band and now UHF-band commensal observing at a significant operational scale. This approach is particularly important for next-generation surveys where large sky coverage and high data fidelity must be achieved simultaneously.

\begin{figure*}
    \centering
    \includegraphics[width=1\linewidth]{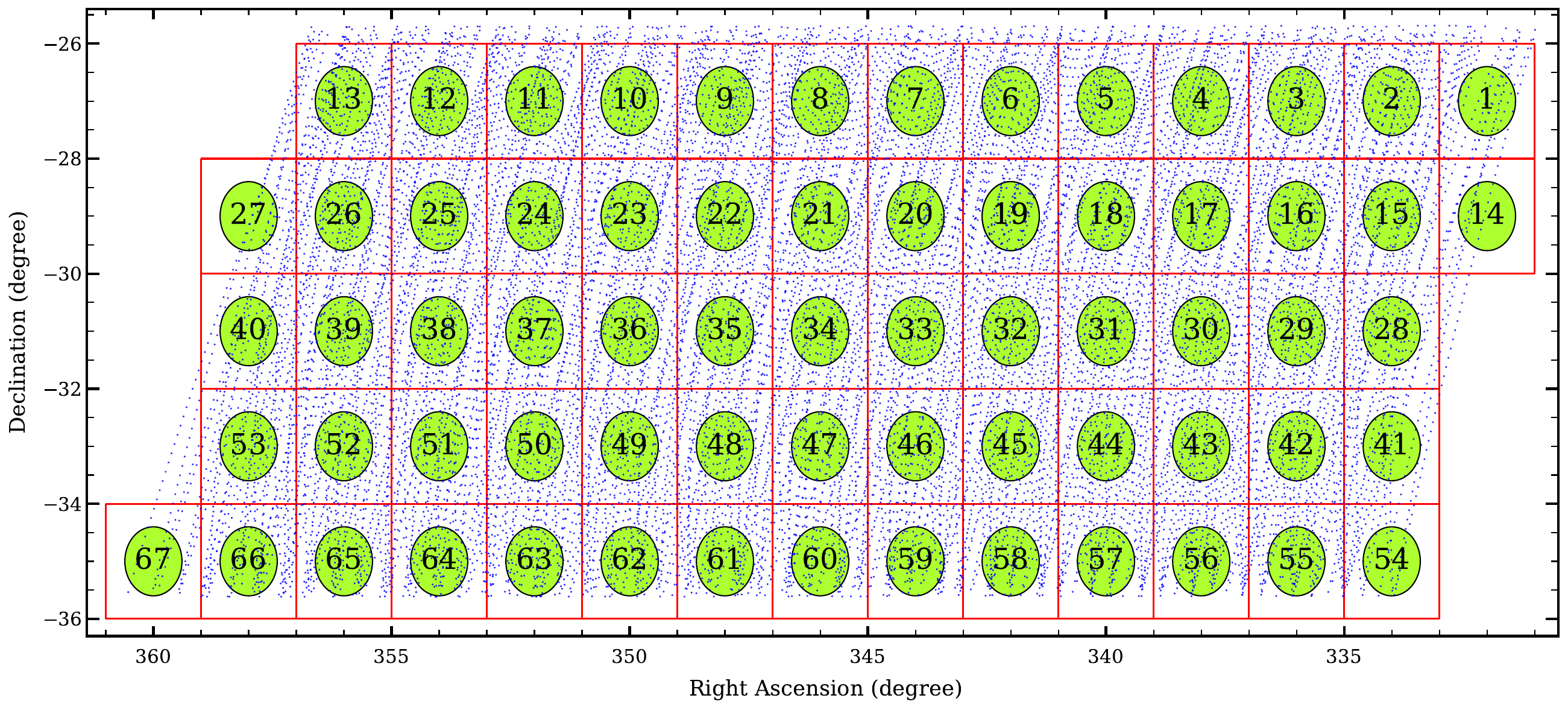}
    \vskip-8pt
    \caption{Sky footprint of the MeerKLASS L-band observing blocks included in this data release. Each blue point marks the pointing centre of the 2\,sec snapshot visibility. This footprint overlaps with the $\text{KiDS-DR5}$ \citep[see][]{KiDS_DR5} and upcoming $\text{DESI}$-DR11. This sky area is divided into $2.15\degr \times 2.15\degr$ tiles with $0.075\degr$ overlap between adjacent tiles. Each square represents a tile, and the tile ID is indicated at its centre. The layout ensures full coverage of the $\sim$268\,deg$^2$ field.}
    \label{fig:alltiles}
\end{figure*}

In the southern sky, wide-area continuum radio surveys have been primarily driven by a combination of key projects. At low frequencies, the GaLactic and Extragalactic All-sky MWA survey \citep[GLEAM;][]{Wayth_2015} and its extension GLEAM-X \citep[][]{Hurley-Walker_2022} map the sky below $+30\degr$ in the frequency range 72–231\,MHz. Around 1\,GHz, the Sydney University Molonglo Sky Survey \citep[SUMSS;][]{Bock_1999,Mauch_2003} and the Rapid ASKAP Continuum Survey at low frequency \citep[RACS-low;][]{Hale_RACS_LOW} deliver wide-area coverage of the southern hemisphere, and the NRAO VLA Sky Survey \citep[NVSS;][]{Condon_1998} covers the sky north of $-40\degr$ at 1.4\,GHz. Within the same ASKAP programme, RACS-mid \citep[][]{Duchesne_RACS_MID_DR1,Duchesne_RACS_MID_DR2} and RACS-high \citep[][]{Duchesne_RACS_HIGH} extend this to higher frequencies, and the Evolutionary Map of the Universe survey \citep[EMU;][]{Norris_2021_EMU,Hopkins_2025_EMU} is targeting a much deeper 900\,MHz survey over most of the southern sky. In the north, the LOFAR Two-metre Sky Survey \citep[LoTSS;][]{LoTSS} provides $\sim 6\arcsec$ imaging at 120–168\,MHz, while the TIFR GMRT Sky Survey \citep[TGSS;][]{Intema_2017_TGSS} covers most of the sky at 150\,MHz with lower resolution.

At higher frequencies, the VLA Sky Survey \citep[VLASS;][]{Lacy_2020} operates at around 3\,GHz, while the Australia Telescope 20\,GHz Survey \citep[AT20G;][]{Murphy_2010} provides complementary coverage at 20 GHz. Together, these data give a broad multi-frequency view of the radio sky. Combining them with MeerKLASS is important for statistical work on galaxy populations, for finding rare or extreme objects, and for studying nearby radio galaxies and spatially resolved star-forming systems. Furthermore, radio observations spanning a broad frequency range play a critical role in the spectral modelling of radio sources, enabling deeper insights into their physical properties and emission mechanisms \citep[e.g.,][]{Galvin_2018,Sinha_2023,Ballieux_2024}.

In the context of the evolving radio sky, the MeerKLASS L-band interferometric survey occupies a unique niche. It offers a balance between the broad sky coverage of surveys like RACS and the depth of deep-field campaigns such as the MeerKAT International GHz Tiered Extragalactic Exploration survey \citep[MIGHTEE;][]{Jarvis_2016, Heywood_2022}.
A second useful MeerKAT comparison survey is the MeerKAT Absorption Line Survey \citep[MALS;][]{MALS_Gupta_2016, MALs_Deka_2024, MALS2_Wagenveld_2024}, it also delivers wideband continuum catalogues from hundreds of non-contiguous L-band pointings centred on bright radio sources. 
Unlike MIGHTEE, which targets $\sim1\,\umu \rm Jy\,beam^{-1}$ sensitivity over a few small, well-studied fields, MeerKLASS provides vastly broader sky coverage while achieving better surface brightness sensitivity and image fidelity than RACS. Compared to MALS, MeerKLASS trades targeted, patchy sampling for uniform wide-area mapping, which is better suited to population-level statistics with fewer position-dependent selection effects. This combination makes MeerKLASS ideally suited for statistically robust source-population studies of in-band spectral indices and spectral curvature mapping. These measurements are critical for classifying AGN and SFGs and for synergy with both low- and high-frequency radio data. Furthermore, the large area and multi-pass strategy provides valuable data for rare sources such as slow radio transients and varying radio sources.

The dataset presented here comprises the first public data release (DR1) of the MeerKLASS L-band interferometric survey. It is derived from eight independent observing blocks (Totalling approximately 13.5 hours of usable time) that cover $\sim 268\,\text{deg}^2$ of the sky. Despite its limited scope, this initial release provides high-fidelity continuum images and a robust catalogue containing 34,874 unique radio sources. These results serve to validate our complex OTF data processing and cataloging pipelines, and already enable several lines of stand-alone science, including source counts and multi-frequency comparisons. 

The structure of this paper is as follows: \autoref{sec:lband-survey} provides an overview of the observations used for this DR1; \autoref{sec:dataprocessing} details the data processing, calibration and crucial OTF phase correction required for fast scanning data, then outlines the  visibility-domain imaging workflow (\autoref{sec:imaging}), describes the validation of the resulting images (\autoref{sec:quality}), presents noise maps and synthesized beam characteristics (\autoref{sec:cataloguegeneration}) and describes the source extraction, catalogue generation, and validation. We then present comparisons with previous radio surveys in \autoref{sec:comparison}, including the spectral index (\autoref{sec:specindex}) and source counts (\autoref{sec:sourcecounts}). We outline the released data products in \autoref{sec:public_release}  
and provide a summary of our findings and future plans in \autoref{sec:Conclusions}.

\section{MeerKLASS L-Band Survey}
\label{sec:lband-survey}

\subsection{Introduction to MeerKLASS}
The MeerKLASS project is conducting a systematic survey of large regions of the extragalactic sky, including regions in both the northern and southern extragalactic cap regions. This target sky area is characterized by large overlaps with several prominent wide-field optical and near-infrared galaxy surveys, a synergy that is expected to yield a highly valuable, multi-wavelength dataset for addressing key questions in extragalactic astronomy.

For this study we are using L-band observations obtained with MeerKAT in fast scanning mode. Continuum images covering the entire MeerKLASS survey area are being produced using the OTF imaging method. Each correlator integration utilizes the full instantaneous bandwidth of the MeerKAT telescope, which is divided into 4096 frequency channels.  The data are obtained with 2\,s integrations. 

\subsection{Survey description}
\label{sec:survey}

For the MeerKLASS survey, the MeerKAT telescope slews back and forth in azimuth at fixed elevation with a $\sim10\degr$ throw at a rate of $\sim10\,\arcmin$/sec, producing repeatable and  stable beam tracks. This strategy is driven by the commensal single-dish intensity mapping part of the survey \citep{Wang_2021,Cunnington_2023}, but it is also beneficial for high-fidelity interferometric mosaicking and continuum image reconstruction. Fixed-elevation scanning reduces variations in system temperature associated with ground spill and atmospheric effects, while the back-and-forth pattern ensures uniform sky coverage. Crucially, any given sky position is observed multiple times as the primary beam sweeps past with many different nearby pointing centres; when these overlapping measurements are combined in the mosaic, residual primary-beam asymmetries are effectively averaged down, yielding a smoother, more nearly symmetric effective beam and reducing beam-related imaging artefacts. This efficient scanning approach yields a nominal survey speed of $\sim$\,150\,deg$^2$\,hr$^{-1}$. with interferometric visibilities recorded every 2\,sec to produce a manageable data rate. During these scan observations the correlator phase center is fixed at the center of the scan in telescope coordinates; as the Earth rotates, this phase center therefore drifts slowly across the sky over the course of a scan block. To mitigate far-sidelobe contamination and solar/ionospheric systematics, all observations are scheduled during night-time Local Sidereal Times (LSTs).

In MeerKLASS, each target field is typically observed twice per night; corresponding to a `rising' scan (observed as the field ascends in the east) and a `setting' scan (observed as the field descends in the west). In both cases, the telescope executes a periodic slew in azimuth along the same constant-elevation track. Due to the Earth's rotation, these rising and setting passes naturally produce cross-linked scans, yielding a cross-hatched pattern in equatorial coordinates. This cross-linked strategy is key to improving coverage uniformity, suppressing striping artefacts associated with single-direction scanning, and enhancing sensitivity by averaging down direction-dependent systematics. A detailed description of the MeerKLASS observation strategy can be found in \citet{Chatterjee_2025_OTF}. The eight block observations utilised in this paper were acquired between September and October 2021, exclusively in the L-band (856$-$1712\,MHz).  Each block involves about two hours of observations, including a total of 20 minutes on the primary calibrator (1934$-$638) split between the beginning and the end of each observing session. The MeerKLASS L-band DR1 focuses solely on continuum imaging and includes only data blocks corresponding to rising scans, yielding a total of $13.5\,\text{hr}$ of usable target data. The calibrator used for the setting scans is an extended source (Pictor A), making the calibration of the setting scans more challenging.  This choice was driven primarily by the commensal autocorrelation (single-dish) intensity-mapping observations, which require an exceptionally bright calibrator; the trade-off is that the source is extended on interferometric baselines, complicating gain and bandpass calibration for the visibility data. Therefore, we plan to include the combination of rising and setting scans for the full dataset consisting of 41 blocks of observing as part of our upcoming DR2 at the end of 2026.

\begin{figure}
\centering
    \includegraphics[width=1\linewidth]{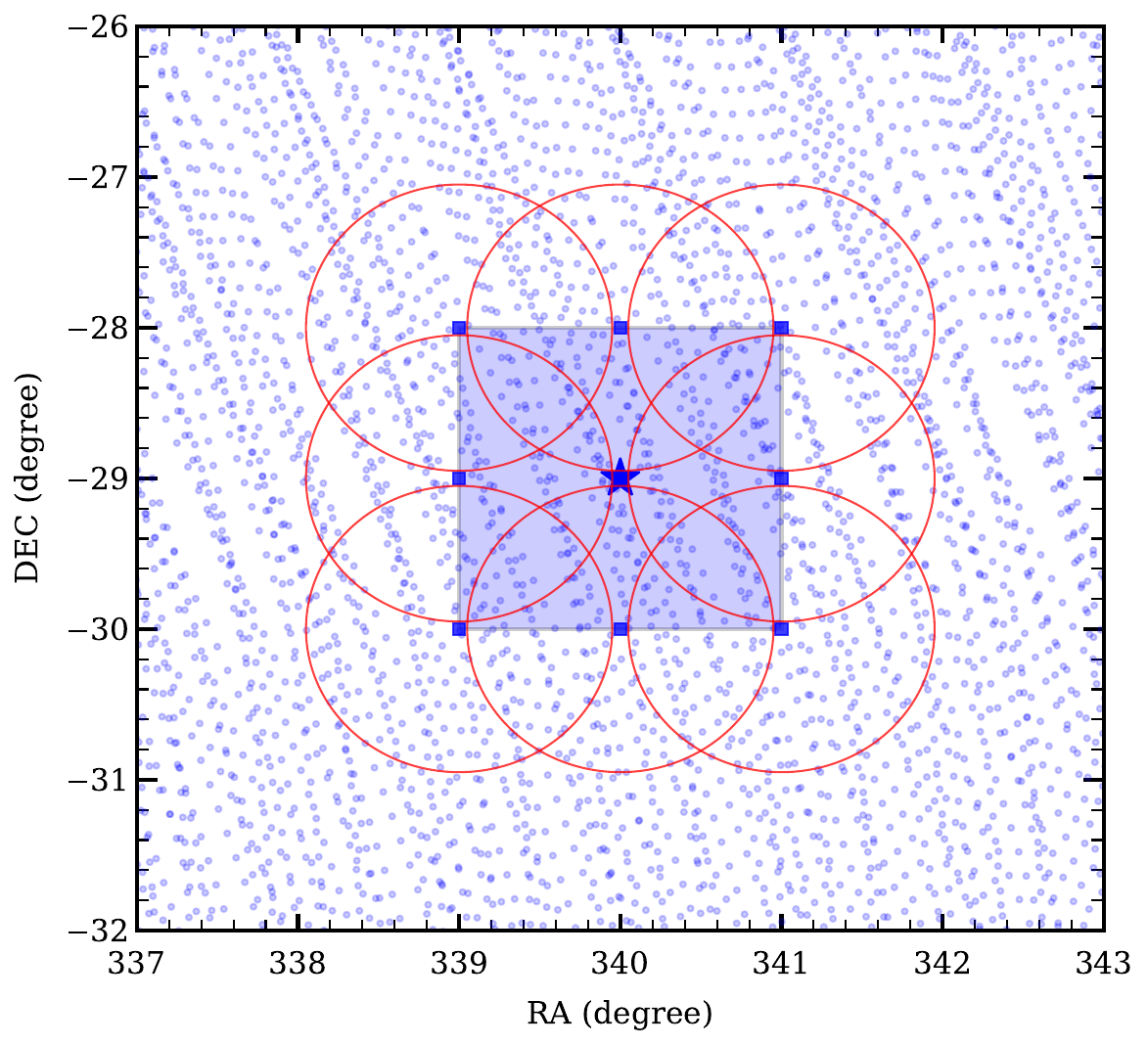}
    \vskip-8pt
    \caption{Illustration of snapshot selection for imaging a single MeerKLASS L-band tile. The blue points show all snapshot pointing centres in the field. The shaded square marks the $2\degr\times2\degr$ region around the chosen tile centre (Blue star). Red circles indicate the $10\,\text{dB}$ primary-beam attenuation contours at 1284\,MHz for the snapshots whose centres fall within and around this region; these snapshots are jointly imaged with \texttt{DDFacet} to form the final tile image.}
\label{fig:snapshot_selection}
\end{figure}

\begin{figure*}
    \centering
    \vskip-0.05in
    \subfloat[][MeerKLASS L-band Stokes I mosaic Image]{
    \includegraphics[width=1\linewidth]{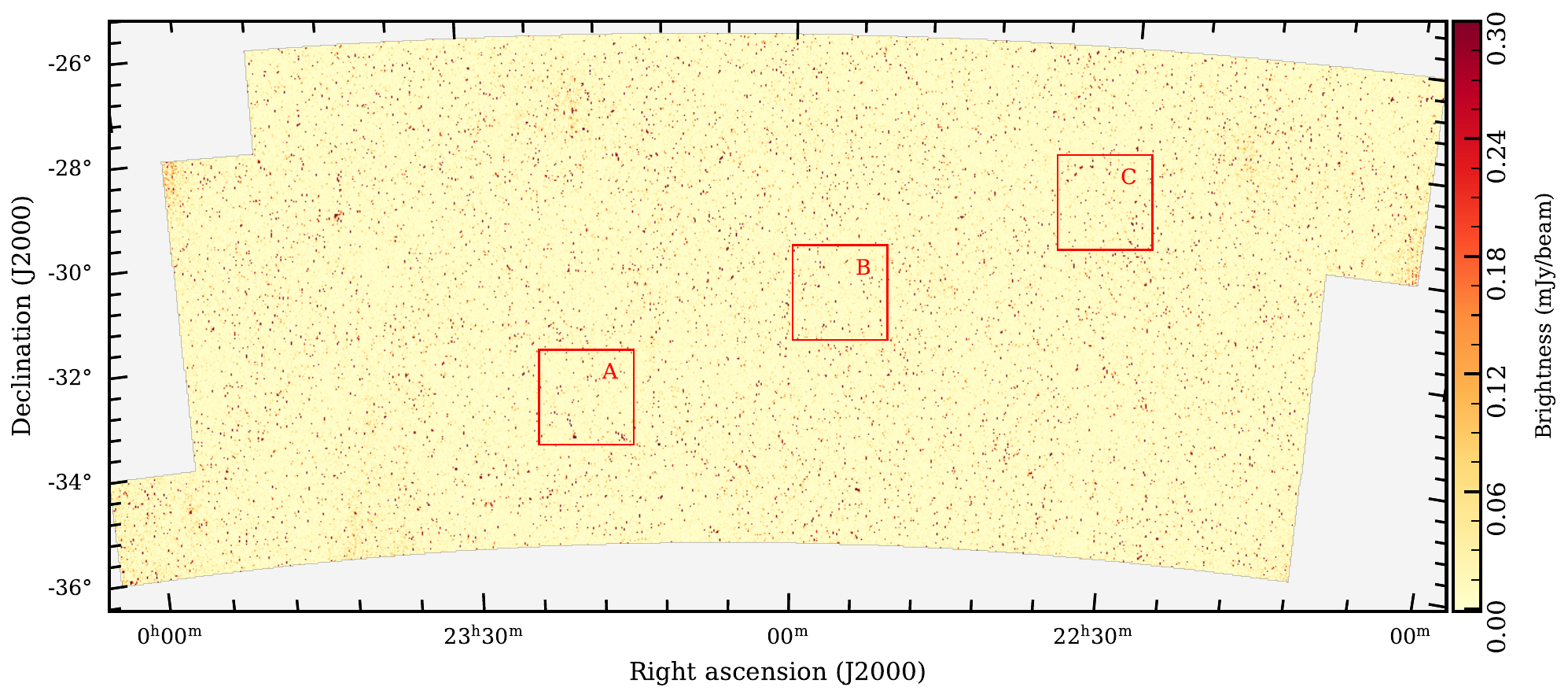}
    \label{subfig:mosaicfull}}
    \vspace{-0.10in}
    \subfloat[][Box A: Tile 50]{ \centering
    \includegraphics[width=0.31\linewidth]{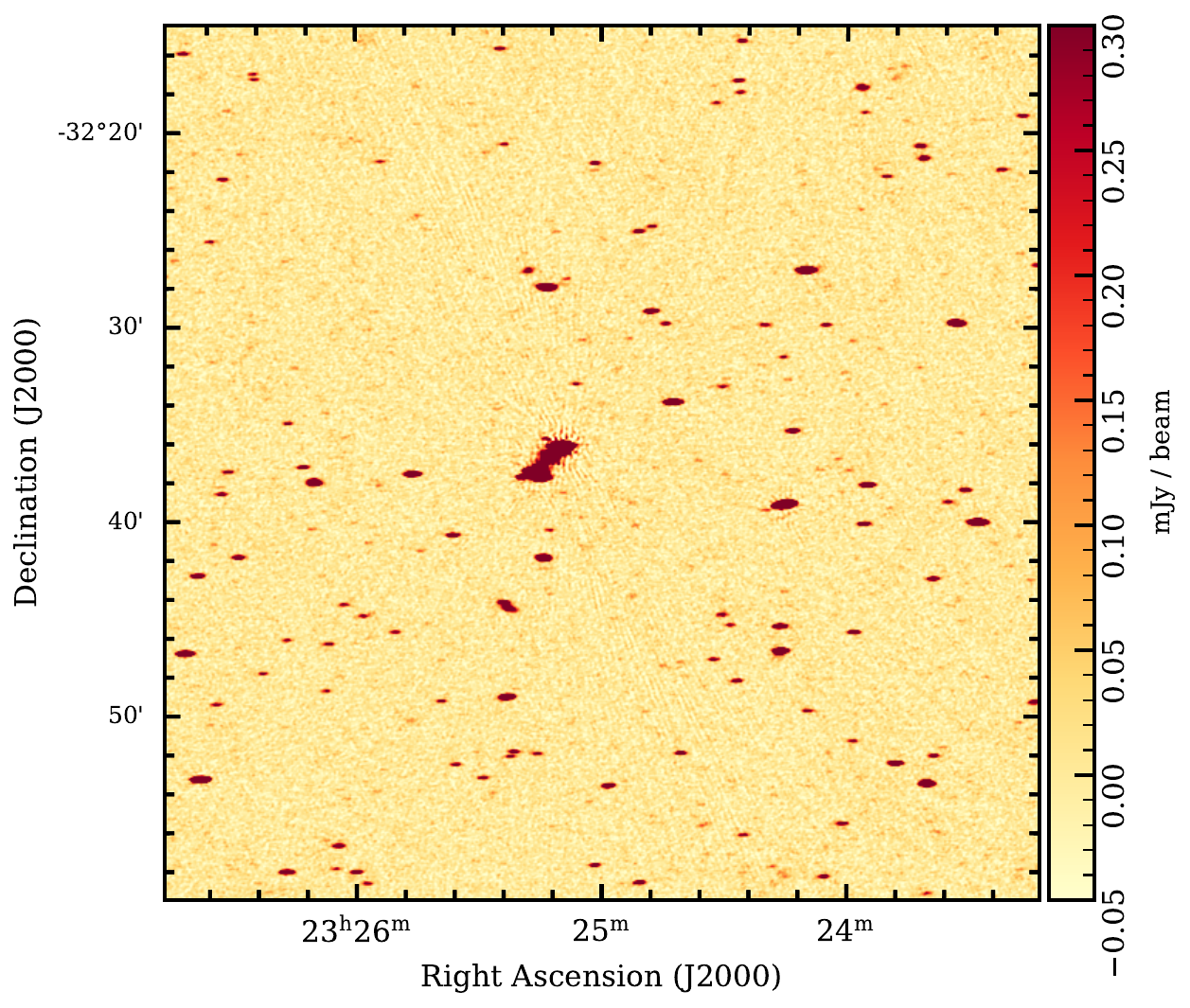}
         %\caption{}
         \label{subfig:tile50}}
    \hspace{1em}
    \subfloat[][Box B: Tile 34]{ \centering
         \includegraphics[width=0.31\linewidth]{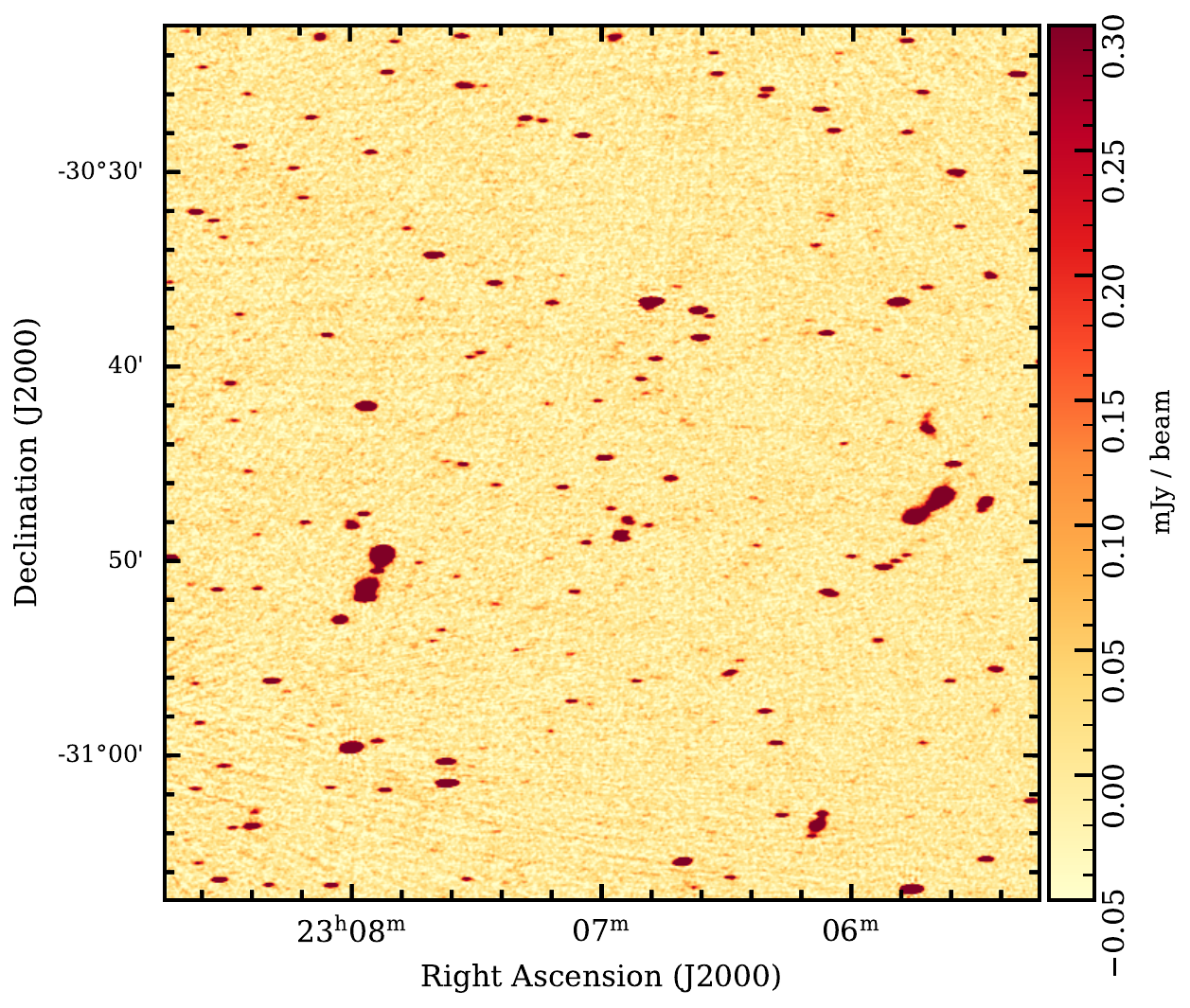}
         %\caption{}
         \label{subfig:tile34}}
    \hspace{1em}
    \subfloat[][Box C: Tile 18]{ \centering
         \includegraphics[width=0.31\linewidth]{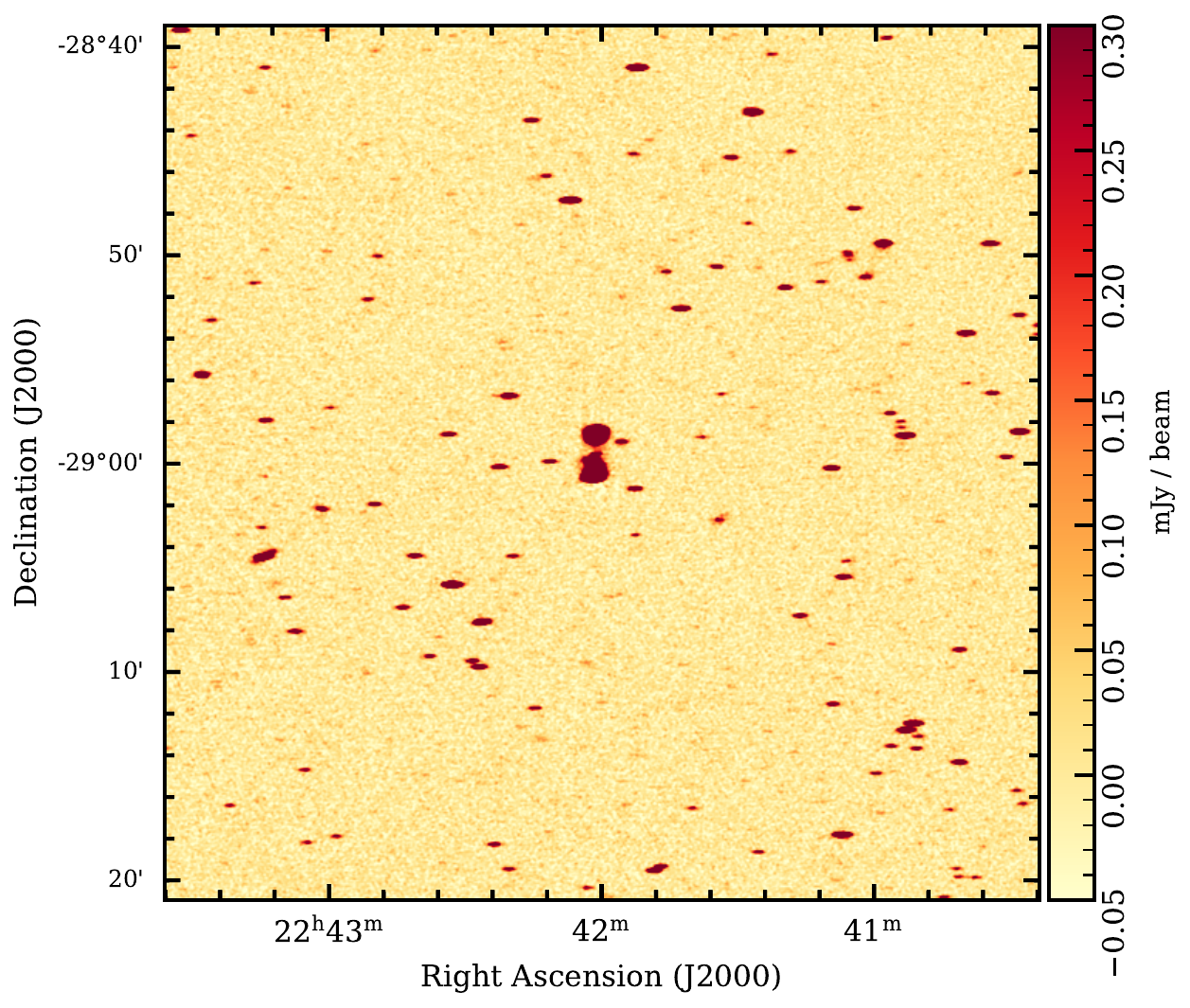}
         %\caption{}
         \label{subfig:tile18}}
         \vskip-4pt
    \caption{Overview of the mosaic and image quality across the MeerKLASS  L-band DR1 survey area used in this work. The top panel shows the full $\sim$268\,deg$^2$ continuum mosaic constructed from 67 tiles (\autoref{fig:alltiles}) at the L-band centred around 1284\,MHz. Three example tiles are highlighted on the mosaic as A, B and C. The zoomed-in cutout of the respective tiles are shown in the lower panels. These illustrate the high dynamic range and low noise floor achieved across a range of sky positions. The selected fields span a variety of declinations and survey depths, and their structure reflects the uniformity and imaging fidelity delivered by the pipeline.  The anisotropy of the synthesized beam is driven by a time smearing effect that has since been overcome with a new OTF-optimized observing mode at MeerKAT.}
    \label{fig:mosaic}
\end{figure*}

These eight rising scan blocks cover an area of approximately 268\,deg$^2$ of sky, spanning a 856\,$\text{MHz}$ bandwidth in the frequency range of 856 to 1712\,MHz. The observations cover Right Ascension $330\degr$ to $360\degr$ and Declination $-26\degr$ to $-36\degr$. All eight blocks overlap within the full area, but due to the scan configuration and spatial layout, the final image sensitivity is not entirely uniform across the footprint. Outer regions near the edge of the footprint have fewer overlapping 2\,s integrations, and therefore exhibit marginally higher noise compared to the central parts where the coverage is denser. A summary of the eight blocks is presented for reference in \autoref{tab:obs_summary}, with their sky locations and scan details illustrated in \autoref{fig:alltiles}.

\section{Data Processing and Calibration}
\label{sec:dataprocessing}
Each observation block is processed using a dedicated calibration and imaging pipeline that converts the raw visibilities into individual calibrated visibilties of 2\,s each with the correlater phase center rotated into the mean pointing direction of the array during the 2\,s integration. The resulting calibrated visibilities are subsequently combined in the visibility domain to generate wide-field continuum images (or mosaics). The processing includes automated flagging, calibration strategy, and quality-control checks to ensure data fidelity before final imaging. A detailed description of the pipeline design, parameter choices, and methodology is provided in \citet{Chatterjee_2025_OTF}, and we summarise here the key steps relevant to this data release.

\subsection{Data flagging, calibration and OTF correction}

\subsubsection{Initial preprocessing and flagging}

Each observing block undergoes an initial preprocessing stage to identify and flag both radio-frequency interference (RFI) and instrument-specific systematics. RFI originating from terrestrial and satellite transmissions (e.g. GNSS, satellite communications, and various navigation signals) is mitigated using the \textsc{\texttt{Tricolour}} \citep[][]{Tricolour_2022} package, integrated within the \texttt{CARACal}\footnote{\hyperlink{https://caracal.readthedocs.io/en/latest/index.html}{https://caracal.readthedocs.io}} \citep[][]{CARACal_2020} pipeline. \textsc{\texttt{Tricolour}} employs a MeerKAT-optimised implementation of the Sum-Threshold algorithm \citep{Offringa_2010} and is applied consistently to both calibrator observations and the target scans. The fraction of flagged data is typically between 40-50 per\,cent across the L-band. This high fraction is primarily driven by strong, persistent RFI that dominates the central frequency range of the band, particularly the portion sensitive to common satellite downlink signals and radio navigation systems, which requires aggressive excision compared to the less-occupied band edges.

Beyond standard RFI excision, a dedicated custom algorithm is employed to detect and flag `rogue' antennas that fail to maintain synchronised motion during the scanning observations. Such antennas may momentarily lag behind or deviate from the nominal pointing trajectory, leading to decorrelated visibilities and imaging artefacts. To mitigate these effects, the algorithm measures the pointing offset of each antenna relative to the median array position, and antennas exhibiting deviations exceeding 0.1\,$\degr$ are flagged either partially or fully, depending on the magnitude and duration of the offset.

\begin{figure*}
    \centering
    \includegraphics[width=1\linewidth]{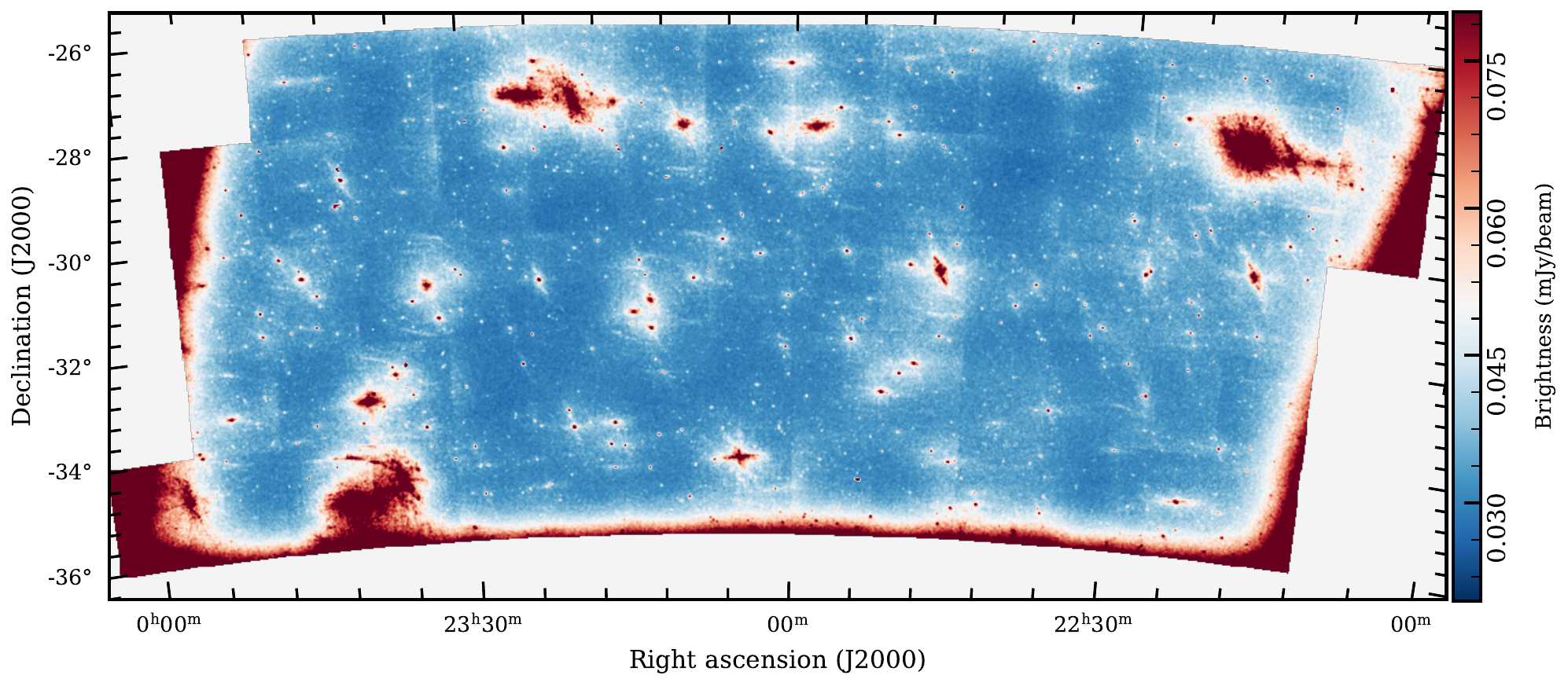}
    \vskip-8pt
    \caption{Estimated background noise across the survey area, derived from the full residual mosaic at 1284\,MHz. The RMS map is computed by applying sigma-clipped statistics within overlapping sliding windows ($100\times100$ pixels, stepped every 50 pixels) directly on the combined residual image. The map reveals significant spatial variation in noise level due to differences in integration time near the survey edge and residual imaging artefacts near bright sources.}
    \label{fig:rms_map}
\end{figure*}

\subsubsection{Calibration}

Calibration of each observing block is also carried out using the automated \texttt{CARACal} pipeline, configured to process each block independently as outlined in \autoref{sec:lband-survey}. The pipeline includes a standard sequence of model definition, primary calibration (bandpass and flux), and time-dependent complex gain calibration, all optimised for the scan-mode observations. The absolute flux scale is tied to standard models for the primary calibrator (J1934$-$638), and the derived gain solutions are interpolated across the target scans to correct for instrumental and atmospheric variations.

The calibration procedure follows the standard \texttt{KGBAKGB} ordering for the primary calibrator, where \texttt{K} represents delay solutions, \texttt{G} complex gain calibration, \texttt{B} bandpass, and \texttt{A} leakage corrections. Bandpass calibration is performed using the full scan duration ($\texttt{solint} = \texttt{inf}$) to ensure high SNR solutions and spectral stability across the band. Most calibration terms are solved over the full scan duration ($\texttt{solint} = \texttt{inf}$), while the final gain calibration is performed at 60\,s intervals to capture short-term temporal variations. The gain solutions are computed using a \texttt{combine} strategy that allows per-scan flexibility and reuses solutions where appropriate to improve efficiency. 

The final calibration tables are then applied to the science data through the \texttt{apply\_cal} step within \texttt{CARACal}. This step applies the derived gain ($\texttt{G}$), bandpass ($\texttt{B}$), and cross-hand ($\texttt{A}$) corrections to the target visibilities.  This approach enables each observing block to be processed independently, ensuring uniformity and reproducibility across the full dataset.

\subsubsection{OTF phase center correction}

In the MeerKLASS scanning survey, the MeerKAT correlator maintains a fixed delay centre at a constant azimuth and elevation throughout each scan. This fixed centre is typically located near the midpoint of the azimuthal sweep. As the antennas continuously slew across the sky, the instantaneous pointing centre, the centroid of the array's primary beam response moves relative to this fixed delay centre. When expressed in equatorial coordinates, this motion introduces a time-dependent offset between the correlator phase centre and the true pointing position of the telescope beam, resulting in a time-varying phase term in the visibilities due to the changing geometric delay.

This time-dependent phase behaviour has two main consequences for the visibility data. First, a fringe rotation is applied so that the visibilities are referenced to the array pointing centre (i.e. the centre of the reconstructed map), corresponding to the mean pointing during the 2\,s correlation integration. Second, because the correlator does not compensate for this continuous motion during the integration period (unlike in a standard tracking observation), the visibility phase changes within each 2\,s integration.  This intra-integration phase variation leads to a reduction in the coherence of the recorded signal, producing amplitude loss that depends on the visibility fringe rate. The impact is strongest on long baselines and toward the high-frequency end of the L-band, where it can produce image smearing and astrometric offsets.

To mitigate the first of these effects, a post-correlation correction is applied using the \texttt{CHGCENTRE}\footnote{\hyperlink{https://wsclean.readthedocs.io/en/latest/chgcentre.html}{https://wsclean.readthedocs.io/en/latest/chgcentre.html}} task from the \textsc{\texttt{WSClean}} \citep{WSclean_2014} package. This task computes and applies the required phase rotation for each visibility, shifting the effective phase centre to the average pointing centre of the antennas during the 2\,s integration. The pointing centre information is extracted per integration from the metadata of a designated reference antenna (typically $\text{m}008$ or $\text{m}009$). This correction is applied at the native 2\,s time resolution, ensuring minimal residual phase errors and preserving astrometric accuracy across the wide field of view.

The second effect, partial coherence loss within each integration is addressed during imaging, as described in detail in \citet[][]{Chatterjee_2025_OTF}. In brief, during imaging the $\texttt{DDFacet}$ algorithm models the frequency- and time-dependent smearing by calculating an effective point-spread function (PSF) that explicitly incorporates this fringe-rate-dependent attenuation. This PSF correction is applied self-consistently throughout the deconvolution process, thereby preserving flux accuracy and image fidelity across the wide-field mosaics.

With this correction, this so-called time smearing leads to an increase in the size of the synthesized beam along the RA direction, resulting in significantly anisotropic synthesized beams in the data used for DR1. In 2025, a new OTF-optimized scanning mode was adopted by MeerKAT, which removes this time-smearing by tracking the correlator phase centre during a single scan and then moving it periodically to keep the correlator phase centre near the centre of the scan sweep as the block observation proceeds.

\subsection{Imaging and calibration workflow}
\label{sec:imaging}
Given the scanning mode of the MeerKLASS observations, a combined $\sim$100\,minute target scan produces approximately 2,800 individual 2\,s visibility datasets. A straightforward approach would be to image each short snapshot independently and subsequently mosaic the results in the image plane, and this is the approach that we took with our prototype pipeline \citep{Rozgonyi-2022eas..conf.1215R,Rozgonyi-2025ASPC..538..336R}. However, this strategy is computationally inefficient, as it requires managing thousands of small images and performing large-scale mosaicking operations. Moreover, the brief integration time of each snapshot results in a relatively low SNR, which not only limits the imaging fidelity and dynamic range of the individual snapshots, but also strongly restricts the feasibility of self-calibration at the snapshot level.

To overcome these limitations, we adopt a ``visibility-domain mosaicking approach'' implemented through a customised version of the $\texttt{DDFacet}$ imaging framework \citep{Tasse_2018}, which is adapted specifically for MeerKLASS OTF corrected visibilities or measurement sets. This method performs joint imaging and deconvolution of multiple overlapping measurement sets directly from the visibilities, ensuring consistent flux calibration and a seamless reconstruction of the sky across the survey footprint. Although explicit direction-dependent calibration is not yet applied, the algorithm incorporates the spatial variation of the primary beam during image formation, thereby enhancing the overall fidelity and uniformity of the resulting mosaics. This visibility-domain strategy allows for the efficient processing of the large L-band survey footprint, which is achieved by partitioning the total survey area into smaller, partially overlapping tiles. 

\subsubsection{Tiles and visibility selection criteria}
\label{sec:tiles}

The sky coverage of the MeerKLASS L-band observations is illustrated in \autoref{fig:alltiles}, where each blue dot represents a 2\,s snapshot visibility. To produce images of this area, we divide the full L-band footprint into 67 distinct fields, or tiles. 

For each tile, a central coordinate is defined and a 2$\degr \times$2$\degr$ region is selected around it. The primary beam attenuation at the $10\,\text{dB}$ level is then computed as circular contours centred on each snapshot pointing, evaluated at the corners and midpoints of the defined region. The $10\,\text{dB}$ level is calculated at 1284\,MHz, the central observing frequency. All snapshot observations whose pointing centres fall within these attenuation boundaries are selected and jointly imaged using the $\texttt{DDFacet}$ software package. Illustration of snapshot selection for imaging a single MeerKLASS L-band tile is shown in \autoref{fig:snapshot_selection}. This approach ensures that all relevant visibilities contributing significant sensitivity to each tile are included in the imaging process. The initial input visibility selection typically covers an area of up to $3.8\degr \times 3.8\degr$. By selecting visibilities such that the central target 2$\degr \times$2$\degr$ region is well-sampled, sources outside the central cutout are properly deconvolved, thereby minimising side-lobe contamination and ensuring a uniform noise floor across the final image.

From the resulting wide-field image, the central $2.15\degr \times 2.15\degr$ region, centred on the tile coordinate, is extracted as the final science image. The overlap between adjacent tiles ensures smooth transitions and consistent noise properties across the full L-band mosaic, and it helps avoid tile edge effects for extended sources located near the edges of a tile.

\begin{figure}
    \centering
    \includegraphics[width=1\linewidth]{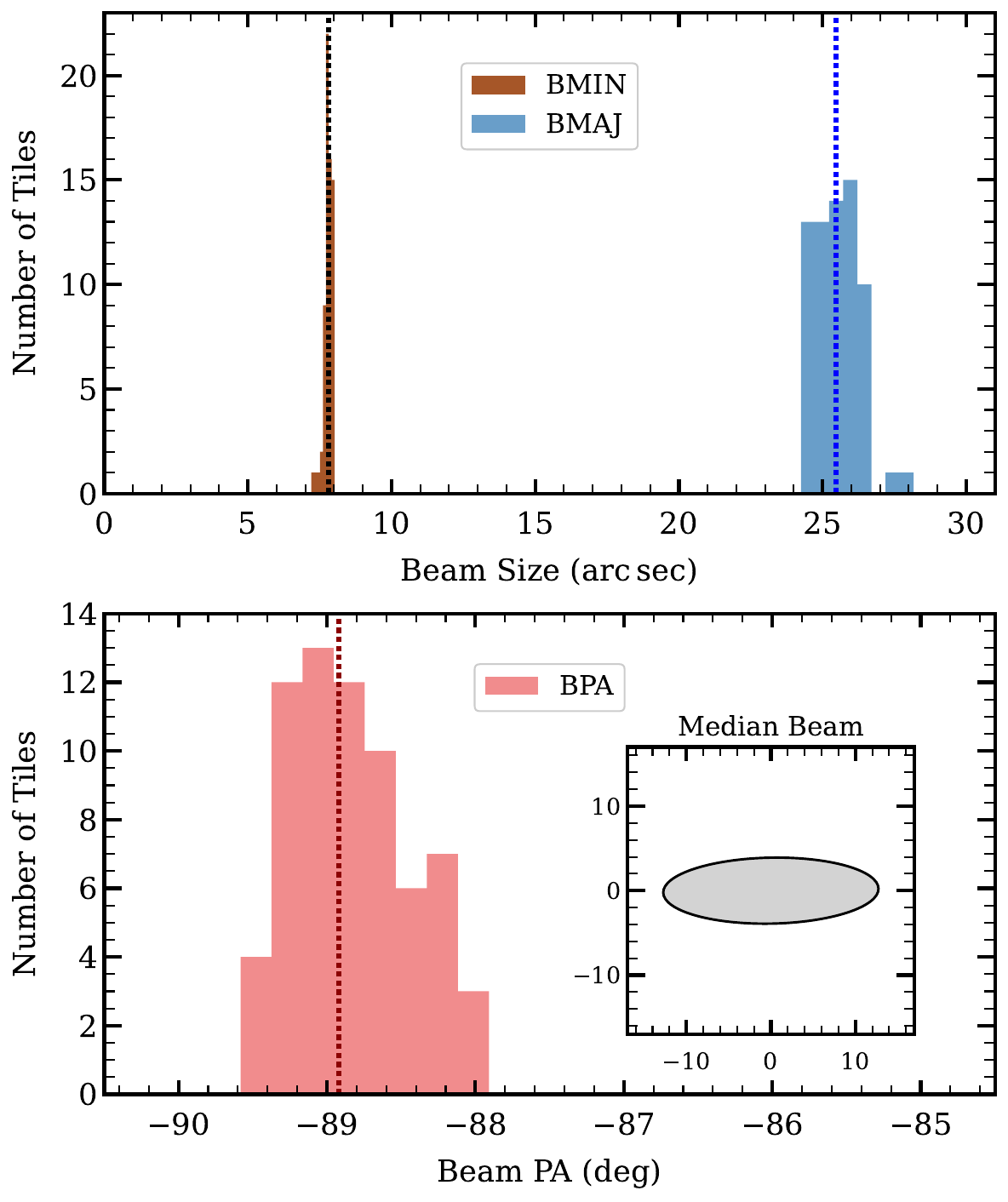}
    \vskip-8pt
    \caption{ Histogram of restoring beam sizes across all 67 tiles. \textit{Top:} The distributions of \texttt{BMAJ} (major axis) and \texttt{BMIN} (minor axis) are tightly clustered around median values of $\sim25.5\arcsec$ and $\sim7.8\arcsec$ respectively (dashed vertical lines), indicating uniform angular resolution over the full survey footprint. \textit{Bottom:} The distribution of \texttt{BPA} (beam position angle) of the major axis clustered around median value of approximately $-88.9\degr$. Also, a median beam is shown in the right panel for reference.}
    \label{fig:beam}
\end{figure}

\subsubsection{Imaging}

The visibilities selected for each tile are subsequently imaged using the \texttt{DDFacet} package. We utilise a facet-based approach, which divides the imaging field into many smaller regions (facets), allowing the visibilities to be locally gridded and deconvolved in each direction before being combined into a single, wide-field image. 
For the L-band, we typically used a $10,240 \times 10,240$ pixel image size, with a pixel scale of approximately 1.5$\arcsec$. The image is deconvolved using the \textsc{\texttt{SSD2}} algorithm, employing Briggs weighting with a robust parameter of 0.

The imaging process is carried out in multiple stages. The initial round involves an automated thresholding procedure (\textsc{\texttt{auto-masking}}) to generate \textsc{\texttt{CLEAN}} masks around compact, high SNR sources within each facet. A subsequent deeper deconvolution stage then follows, utilizing externally generated masks derived from smoothed residual maps and model-based source detection from the preliminary imaging run. This second stage is critical for recovering extended and faint emission, which are not well captured in the first pass. 

Following the deeper imaging pass, the \texttt{killMS} package \citep[e.g.,][]{Tasse_2023} is applied to perform direction-independent self-calibration. Complex, per-antenna, full-Stokes gain solutions are derived at 60\,s intervals across the observed band, leveraging the $\textsc{\texttt{CLEAN}}$ component model generated from the second imaging run. The resulting solutions 
are then smoothed in both the time and frequency domains before being uniformly applied to all snapshot visibilities.

A final imaging iteration is executed using fixed $\textsc{\texttt{CLEAN}}$ masks and the restored source models to enhance PSF stability and suppress residual calibration artefacts. Imaging is conducted in seven frequency sub-bands ($\texttt{Freq-NBand 7}$), producing a multi-frequency cube. These frequency-resolved products provide valuable inputs for subsequent spectral and continuum analysis, enabling accurate measurement of spectral indices and spectral curvature across the field.

The imaging process described above creates, for each tile, a model image, a restored image, a residual image, and beam information in standard \texttt{FITS} format, as well as the seven-band image cube for spectral analysis. The primary data products for the MeerKLASS-OTF L-band release are continuum total intensity images and sub-band images. The total intensity MeerKLASS-OTF mosaic image for L-band is shown in \autoref{fig:mosaic}.

\subsection{Image quality assessment}
\label{sec:quality}
Prior to source extraction and catalogue construction, we check the quality of the final restored images. This includes visual checks, statistical analysis of image noise, and examination of the PSF and residual artefacts. These checks help us confirm our approach and set expectations for catalogue reliability and completeness. 

\begin{figure*}
    \centering
    \includegraphics[width=1\linewidth]{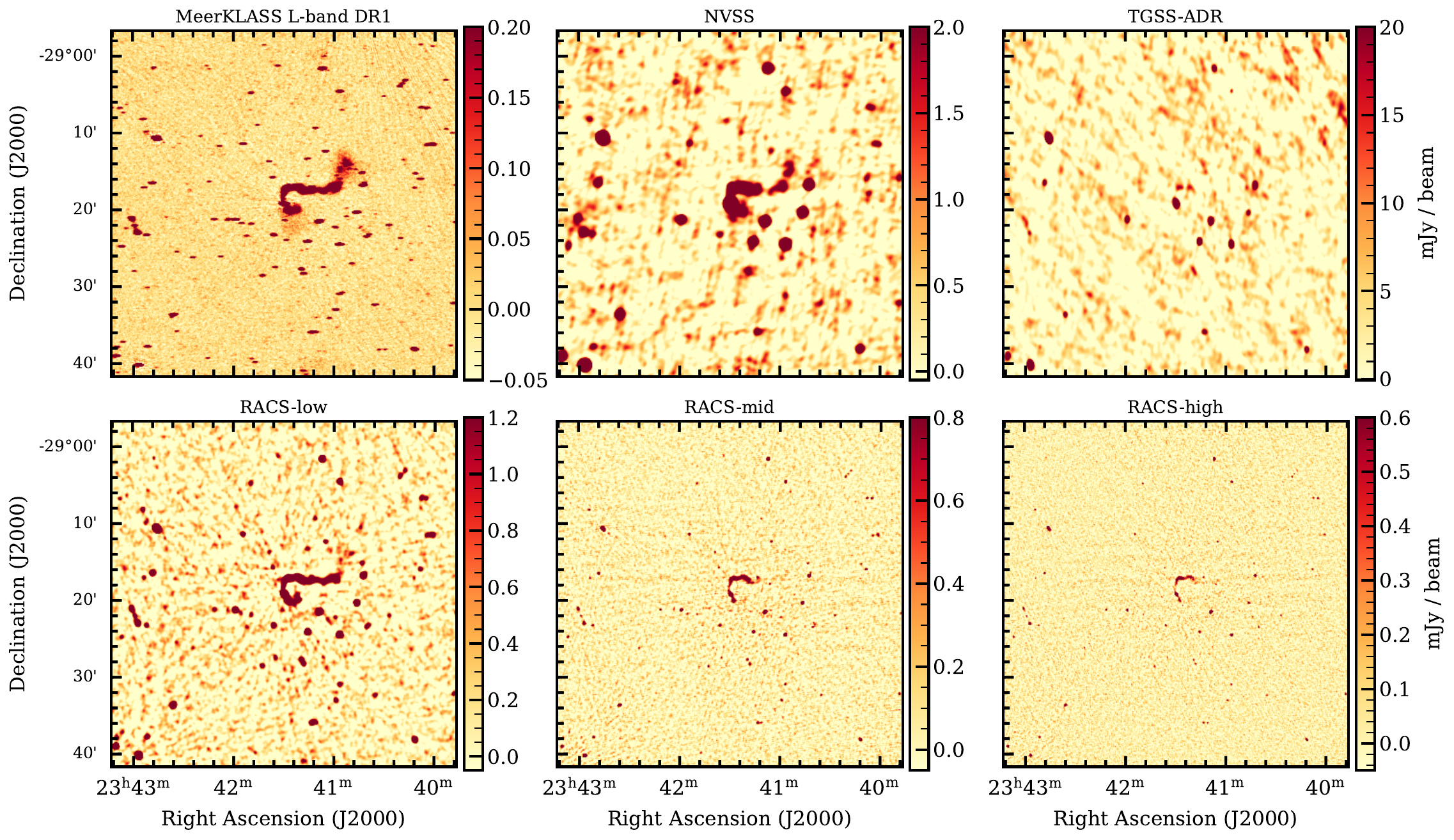}
    \vskip-8pt
    \caption{Comparison of the MeerKLASS L-band image with major radio surveys. Each image is centred around the $\rm J234129.7 -291915.9$ source. Top panel: MeerKLASS L-band DR1 (Freq: 1284\,MHz; Beam: $25.5\arcsec \times 7.8\arcsec$) image with NVSS (1400\,MHz; $45\arcsec$) and TGSS-ADR (150\,MHz; $25\arcsec$) overlaid for the same sky region. Lower panels: RACS-low (887.5\,MHz; $25\arcsec$), RACS-mid (1367.5\,MHz; $11.4\arcsec \times 9.3\arcsec$), and RACS-high (1655.5\,MHz; $7.4\arcsec \times 6.2\arcsec$) images of the same field. Flux-density scales differ between panels and are set by each survey’s sensitivity. The MeerKLASS L-band DR1 image reveals substantially fainter sources and more extended emission than the other surveys.}
    \label{fig:comparison}
\end{figure*}

\subsubsection{Full mosaic}

A composite mosaic image is constructed from the 67 individual restored images (or tiles) shown in \autoref{fig:mosaic}. We construct the final mosaic with  \textsc{Montage}\footnote{\url{http://montage.ipac.caltech.edu}} \citep[][]{Jacob_2010}, which reprojects each \texttt{FITS} image to a common \texttt{WCS} frame before co-adding them into a seamless large-area product. To ensure visual continuity and suppress edge effects, each tile covers a $2.15\degr \times 2.15\degr$ field, sharing a consistent $1.5\arcsec$ pixel scale and an overlap of approximately $0.075\degr$ with adjacent neighbours. $\textsc{Montage}$ is used in its standard reprojection mode without any background matching or flux rescaling applied, preserving the original image intensities produced by the pipeline.

To assess the fidelity and depth achieved in MeerKLASS DR1, where median RMS sensitivity of $\sim 33\,\umu\text{Jy}\,\text{beam}^{-1}$, we examine Tile 18, which is highlighted as a representative example in the lower-right panel of \autoref{fig:mosaic}. Crucially, Tile 18 is free of any bright radio sources both within and immediately surrounding its field of view, making it an ideal region to assess the true thermal-noise limit of the deep L-band data. The image for Tile 18 demonstrates high quality, showing clean background regions and minimal residual artefacts. It achieves a RMS noise floor of $29\,\umu\text{Jy}\,\text{beam}^{-1}$, consistent with expectations given MeerKAT sensitivity and the effective observing time corresponding to eight L-band observing blocks.

These visual diagnostics confirm that high-fidelity continuum maps covering wide sky regions are produced by the pipeline in the adopted visibility-domain mosaicking and imaging technique.  The robustness of the entire processing pipeline in handling the OTF interferometric data is supported by the absence of strong artefacts or residual sidelobes structure. The more quantitative elements of image quality assessment, such as noise estimates and PSF characterization, follow this visual inspection.

\subsubsection{RMS noise map}

To accurately characterize the sensitivity variation across the MeerKLASS L-band survey footprint, we construct a spatially resolved RMS noise map from the residual images produced during the \texttt{DDFacet} imaging process. Because residual images have had the true radio sources removed during the deconvolution process, they provide a direct measure of the local background fluctuations in each tile, including thermal noise and any remaining calibration or deconvolution residuals.

We utilize \textsc{Montage} to combine the residual images from all 67 tiles into a single mosaic on a common astrometric grid. This allows us to compare pixels across the full survey area. To measure the local RMS, we extract sigma-clipped statistics in a sliding box. A $100\times100$-pixel window (about $2.5'\times2.5'$ at $1.5\arcsec$ resolution) is moved in 50-pixel steps across the mosaic. Within each window, pixels more than $3\sigma$ from the median are iteratively clipped, and the standard deviation of the remaining distribution is adopted as the local RMS. This approach reduces the contamination of side-lobes and poorly subtracted sources and produces a 2D noise map at a coarser spatial scale that tracks large-scale sensitivity changes.

The resulting map is shown in \autoref{fig:rms_map}. The deepest regions correspond to areas with the highest effective integration time from overlapping OTF coverage. As noted in the previous subsection, the deepest tiles (e.g., Tile 18) reach $\sim29\,\umu\mathrm{Jy,beam^{-1}}$, consistent with expectations for MeerKAT L-band given the effective exposure.   Noise increases toward the survey boundaries where the coverage becomes less uniform and primary-beam attenuation is more severe. The map also shows localized ``red patches'' of elevated RMS around very bright and/or extended sources. There, imperfect deconvolution leaves residual PSF sidelobes that increase the local RMS. These patches indicate dynamic-range limitations in the imaging, rather than an intrinsically higher thermal-noise floor. These elevated RMS areas serve as important diagnostics, highlighting regions where local imaging fidelity is reduced and guiding the interpretation of faint source populations. This RMS map plays a key role in assessing survey uniformity, evaluating source detection completeness, and supporting accurate flux uncertainty estimates in the final catalogue. Flux-density uncertainties are additionally corrected for the contribution from time-averaging smearing in the OTF scans; see Appendix~\ref{app:noise_psf} for details.

\begin{table*}
    \centering
    \caption{Residual-based dynamic range for a subset of bright, point-like sources.}
    \begin{tabular}{ccccccc}
        \hline
        \texttt{Source\_Name} & \multicolumn{1}{c}{\texttt{RA}} & \multicolumn{1}{c}{\texttt{DEC}} & \multicolumn{1}{c}{\texttt{Tile\_ID}} & \multicolumn{1}{c}{$S_{\rm peak}$} & \multicolumn{1}{c}{$\sigma_{\rm local }$}  & \multicolumn{1}{c}{Dynamic Range}  \\
        & \multicolumn{1}{c}{(deg)}                                            & \multicolumn{1}{c}{(deg)}  & & \multicolumn{1}{c}{(Jy\,beam$^{-1}$)}   & \multicolumn{1}{c}{(Jy\,beam$^{-1}$)}           & \\
        \hline
        
        MeerKLASS\_L\_DR1 J225600.1-273556.4	&   344.00051	 &  -27.5990	&  7      &	     0.4353	 &   0.0008	  &      520         \\
        
        MeerKLASS\_L\_DR1 J232447.4-271920.3	&   351.19742	 &  -27.3223	&  11     &	     1.2278	 &   0.0102	  &      120         \\
        
        MeerKLASS\_L\_DR1 J221336.7-280346.0	&   333.40293	 &  -28.0628	&  15     &	     0.7214	 &   0.0014	  &      492         \\
        
        MeerKLASS\_L\_DR1 J230305.8-303011.8	&   345.77410	 &  -30.5033	&  34     &	     0.4077	 &   0.0005	  &      755         \\
        
        MeerKLASS\_L\_DR1 J231448.5-313839.8	&   348.70191	 &  -31.6444	&  35     &	     0.5592	 &   0.0012	  &      483         \\
        
        MeerKLASS\_L\_DR1 J231315.1-312504.9	&   348.31294	 &  -31.4180	&  35     &	     0.5418	 &   0.0015	  &      360         \\
        
        MeerKLASS\_L\_DR1 J233505.8-311606.4	&   353.77425	 &  -31.2684	&  38     &	     0.4054	 &   0.0005	  &      761         \\
        
        MeerKLASS\_L\_DR1 J233343.4-305757.0	&   353.43073	 &  -30.9658	&  38     &	     0.8444	 &   0.0016	  &      528         \\
        
        MeerKLASS\_L\_DR1 J224500.2-343030.6	&   341.25091	 &  -34.5085	&  57     &	     0.4077	 &   0.0005	  &      742         \\
        
        MeerKLASS\_L\_DR1 J234145.8-350622.9	&   355.44098	 &  -35.1064	&  65     &	     1.9497	 &   0.0043	  &      454         \\
        \hline
    \end{tabular}
    \label{tab:brightsources}
\end{table*}

To quantify the degree of dynamic-range limitation around bright, point-like sources, we compute a residual-based dynamic range,

\[
\mathrm{DR} \;=\; \frac{S_{\rm peak}}{\sigma_{\rm local}},
\]

where $S_{\rm peak}$ is the peak brightness measured on the restored image, and $\sigma_{\rm local}$ is a sigma-clipped RMS estimated from the residual image in an annulus spanning $2$ to $5$ times the synthesized-beam FWHM. We apply this measurement to the subset of bright, point-like sources to characterise the distribution of $\mathrm{DR}$ across the footprint. Representative examples are listed in Table~\ref{tab:brightsources}, illustrating the spread in $\sigma_{\rm local}$ at fixed $S_{\rm peak}$ and the corresponding range of $\mathrm{DR}$ across different tiles. Low-$\mathrm{DR}$ cases typically occur either (i) in fields affected by residual sidelobes and calibration/deconvolution imperfections near very bright or extended emission, which locally inflate $\sigma_{\rm local}$, or (ii) toward the survey coverage boundaries where non-uniform coverage  increase the local noise. Away from these regions, the images more closely approach the thermal-noise regime.

\subsubsection{Synthesized beam characterisation}

The synthesized beam (or the PSF) sets the image resolution and affects source morphology and completeness. In the MeerKLASS OTF data, the PSF changes across the field are moderate.  They are driven by (i) each tile having a different number of calibrated visibilities, which changes the density and coverage of the visibilities in the $u-v$ plane, and (ii) time-dependent flagging (e.g., RFI removal) introduced variability in the $u-v$ sampling. 

To check beam quality and uniformity, we extract the restoring-beam parameters (major axis (\texttt{BMAJ}), minor axis (\texttt{BMIN}), and position angle (\texttt{BPA}) from the final restored images for all 67 L-band tiles. These are the Gaussian beams fitted to the synthesized PSF at the end of the $\texttt{DDFacet}$ deconvolution.
\autoref{fig:beam} shows the histograms of $\texttt{BMAJ}$ and $\texttt{BMIN}$ over the full footprint. The distributions indicating good uniformity of angular resolution and orientation of the anisotropic PSF. The median \texttt{BMAJ} is $\sim25.5\arcsec$ and the median \texttt{BMIN} is $\sim 7.8\arcsec$. These major axis values are impacted by the time-smearing correction in $\texttt{DDFacet}$ due to the Earth's rotation during the original MeerKLASS observations (since resolved with an OTF-optimized MeerKAT scanning mode). As described in methodology of \citet{Chatterjee_2025_OTF}, this correction compensates for time-averaging during correlation and leads to a broader effective PSF than a static, uniformly weighted L-band observation. The measured beams reported here are therefore the true image resolution of the final maps.

\subsection{Stokes I catalogue generation}
\label{sec:cataloguegeneration}
To detect radio sources and create a catalogue of their properties, we make use of the Python Blob Detection and Source Finder package ($\mathbf{\texttt{PyBDSF}}$; \citealt{Mohan2015}). This tool is applied independently to each image tile of $2.15\degr \times 2.15\degr$. 

\subsubsection{\texttt{PyBDSF} settings}
The general practice of \texttt{PyBDSF} cataloging involves first deriving the RMS noise and sky background maps for an image using a sliding box, after which islands of emission that exceed a user-determined threshold are identified. We run \texttt{PyBDSF} on each image using the detection parameters optimized for the MeerKLASS L-band imaging characteristics, as detailed below:

\begin{verbatim}
bdsf.process_image(image, psf_vary_do=True, 
atrous_do=True, rms_box=(150,50), 
rms_map=True, adaptive_rms_box=True, 
adaptive_thresh=150, rms_box_bright=(20,7), 
thresh_isl=5.0, thresh_pix=7.0)
\end{verbatim}

where {\it image} is the file name of the \texttt{FITS} image for the tile, \texttt{thresh\_isl} is the threshold for detecting an island of emission to then be fit with Gaussian components, and \texttt{thresh\_pix} is the criterion which determines which sources are included within the final source catalogue. Noise estimation is performed using \texttt{rms\_map} with a fixed \texttt{rms\_box}, and the adaptive RMS mode (\texttt{adaptive\_rms\_box}) to avoid detecting artifacts as components/sources around bright sources. Additionally, we also apply\texttt{rms\_box\_bright} to allow finer structure modeling without biasing the global RMS. To improve source characterization in regions with slowly varying PSF and complex backgrounds, we enable wavelet-based detection \texttt{atrous\_do} and allow for spatially varying PSFs by enabling \texttt{psf\_vary\_do}.

Following source detection, the individual $\texttt{srl}$ catalogues generated by $\texttt{PyBDSF}$ for each tile are merged to create a unified source catalogue. For traceability, the originating tile ID, and the beam parameters ($\texttt{BMAJ}, \texttt{BMIN}, \texttt{BPA}$) are recorded as metadata for every entry in the combined catalogue. 

Due to the overlap (typically $\sim 0.075\degr$) between adjacent tiles, a subset of sources is detected in more than one image. We identify and remove these cross-tile duplicates with a simple matching–and–merging step so that the final catalogue contains only unique physical sources. First, we perform a self-match of the concatenated catalogues using the $\texttt{Astropy}$ method $\texttt{SkyCoord.search\_around\_sky}$ with a maximum separation of $1.5\arcsec$ (pixel size).  We retain only links between entries originating from different tiles. The resulting links define connected components (or spatial clusters) among the catalogue entries. Within each connected component, a single representative detection is selected using a data-driven scoring system to prioritize the most robust measurement: Highest Peak Signal-to-Noise Ratio ($\texttt{Peak\_flux}/\texttt{Isl\_rms}$). If there is a tie, the higher $\texttt{Total\_flux}$ is adopted. If there is still a tie, the entry with the smallest angular distance to its original tile centre (as a final tie-breaker, favouring detections measured farther from the tile boundary). All non-selected cross-tile duplicates are flagged and excluded from the final catalogue. The resulting catalogue contains 34,874 unique sources detected over the full MeerKLASS L-band DR1 footprint of approximately $268\,\text{deg}^2$. A detailed description of the catalogue fields is provided in Appendix \ref{app:column_description}.

\subsubsection{Catalogue columns}

Using the combined PyBDSF catalogue, we select a subset of the column information when generating the final catalogue. These columns provide information on: IDs; position; flux density; shape information and other important source information. We present a description of the column information for these two MeerKLASS Stokes I catalogues in \autoref{sec:public_release}.

\section{Comparison with other radio surveys}
\label{sec:comparison}

Having completed the merged catalogue, we now compare our MeerKLASS L-band results with earlier radio surveys across a range of frequencies to validate the source properties derived from our data. We note that we continue to work to improve the images produced in the MeerKLASS survey, and in particular there is now an OTF-optimized scan observing mode at the MeerKAT observatory, and therefore the data products in future data releases  should improve further.

\subsection{Comparison images of extended sources}

We perform a visual comparison of several representative MeerKLASS sources with their counterparts in the  TGSS-ADR1, SUMSS, NVSS, RACS-low, -mid and -high surveys to illustrate differences in angular resolution and sensitivity to diffuse emission. The overlap region suitable for comparison is limited to a narrow declination range ($-36\degr < \delta < -26\degr$), where MeerKLASS, TGSS-ADR1, NVSS and the three RACS bands all provide coverage; SUMSS contributes only for declinations $\delta < -30\degr$.

\autoref{fig:comparison} presents radio images of the giant radio galaxy $\rm J234129.7-291915.9$ from major radio surveys, alongside the MeerKLASS L-band DR1 image. Note that the color scale in each panel is adjusted to provide information close to the noise level to give the best available information about the low surface brightness structures in the jets.  These six panels demonstrate the increased sensitivity of MeerKLASS in comparison to TGSS-ADR1, SUMSS and NVSS, and also the improved angular resolution in comparison to three of the surveys. In particular, the MeerKLASS image reveals continuous emission along the jets and extended, diffuse lobes that are either severely truncated or completely absent in the lower-sensitivity maps. One can see that the overall morphology of this extended source, as defined by MeerKLASS, is not well reproduced by any of the other surveys, with perhaps the exception of RACS-low.

The three RACS bands provide a particularly informative comparison at similar frequencies. RACS-low recovers the overall shape of the jet and both lobes, but the background is noticeably more structured, and the faint outer edges of the lobes are harder to trace into the noise. RACS-mid provides a sharper view of the bright jet spine and central component, yet the lowest surface-brightness emission along the outer bends already starts to fragment and fade. In RACS-high, only the bright inner part of the jets and the core remain prominent, with much of the extended diffuse structure resolved out or suppressed by the higher observing frequency. Taken together, the MeerKLASS and RACS panels highlight that high surface-brightness sensitivity at $\sim 20\arcsec-30\arcsec$  resolution is essential for robustly characterising the full morphology of giant radio galaxies, and that MeerKLASS uniquely adds new information on the low surface brightness outskirts that is not accessible in existing all-sky surveys. Images in these regions will be improved with our future Data release as well as in the future with surveys such as the Evolutionary Map of the Universe \citep[EMU;][]{Norris_2021_EMU, Hopkins_2025_EMU}.

\begin{figure}
    \centering
    \includegraphics[width=1\linewidth]{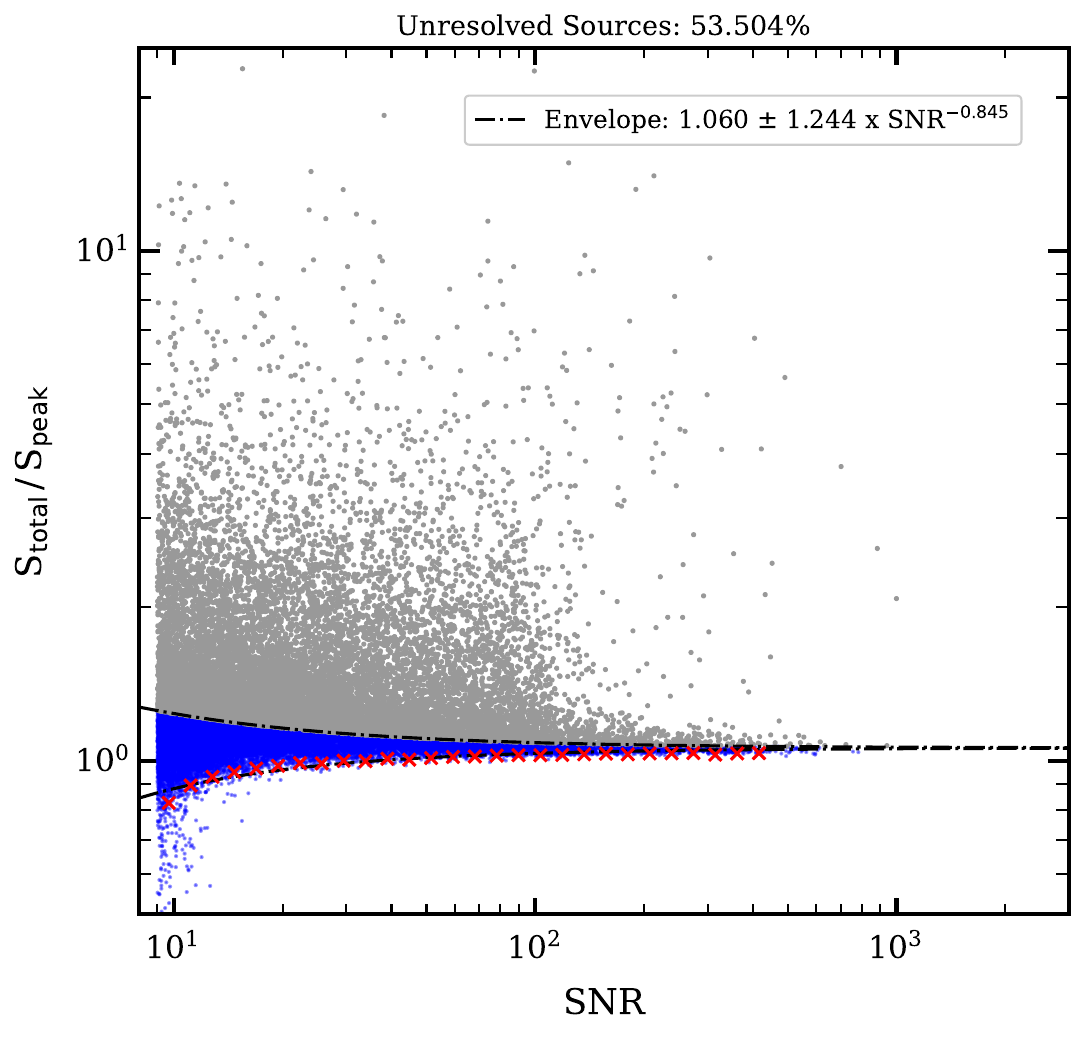}
    \vskip-8pt
    \caption{The ratio of integrated flux density to peak flux density as a function of SNR for single component sources at $\geq 9\sigma$. The lower and upper envelope are drawn as dashed black lines used to define unresolved sources. Blue points mark sources classified as unresolved, while grey points mark sources classified as resolved, based on the envelope criteria described in Section \ref{sec:unresolved}. Red crosses indicate the binned values used to fit the lower envelope.}
    \label{fig:unresolved}
\end{figure}

\subsection{Flux offsets, astrometric offsets}
\label{sec:offsets}

To ensure both flux scale accuracy and astrometric reliability, and to assess how the derived spectral indices align with expectations from the known radio source population, we compare our measurements with several major wide-area radio surveys: NVSS, SUMSS, TGSS-ADR1, RACS-low, -mid, and -high. These surveys span a broad range of observing frequencies, angular resolutions, and sky coverage, with partial overlap in the regions relevant to our study. Given these differences in instrumental characteristics and sensitivity to extended emission, we restrict our comparison to compact, high–signal-to-noise, and isolated sources. This selection minimizes biases introduced by resolution effects or diffuse structure, allowing for a robust assessment of flux consistency, positional accuracy, and spectral behaviour.

\begin{figure*}
     \centering
     \subfloat[]{
         \includegraphics[width=0.31\linewidth]{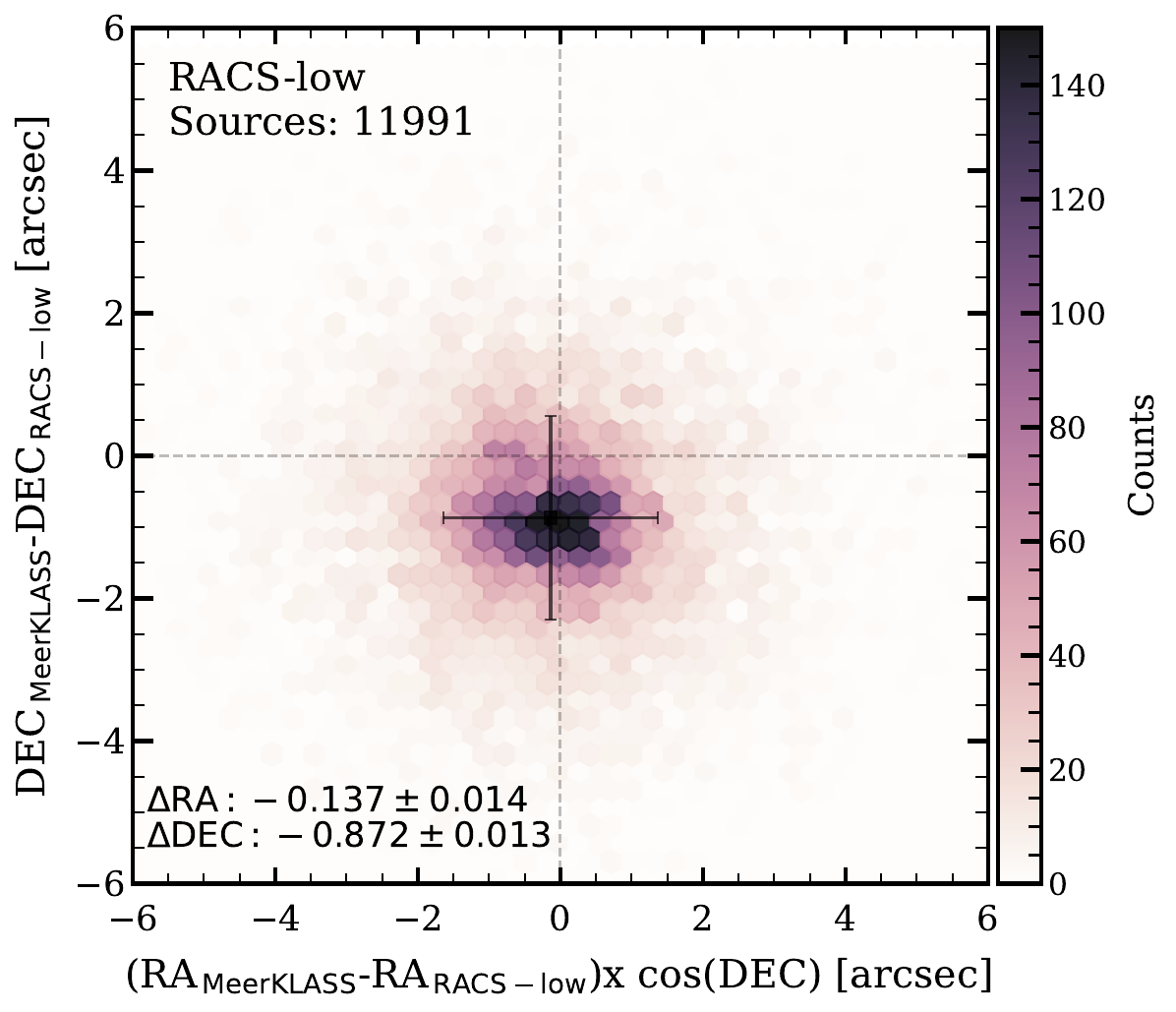}
         %\caption{}
         \label{subfig:astrometrya}}
    \hspace{1em}
    \subfloat[]{ \centering
         \includegraphics[width=0.31\linewidth]{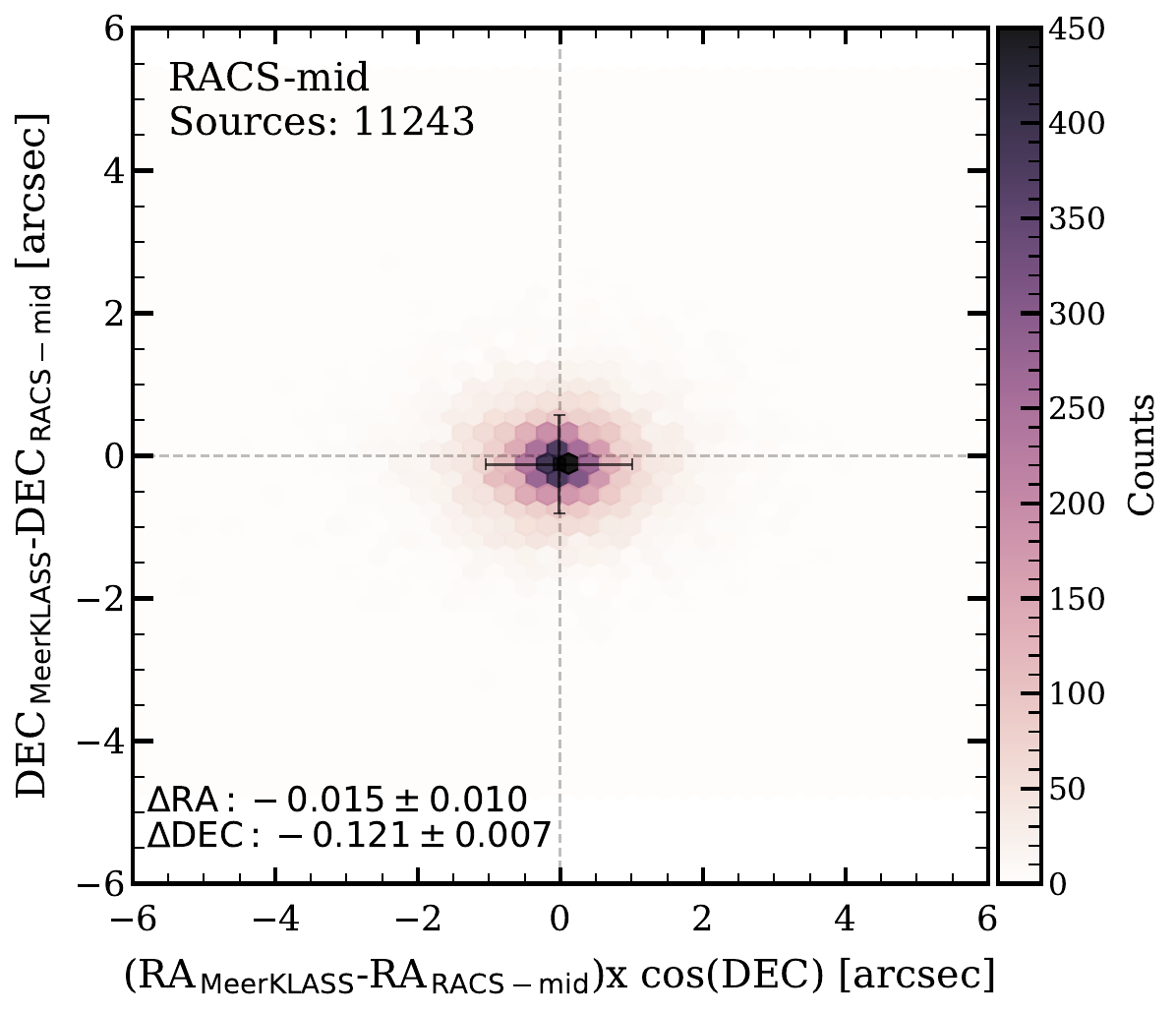}
         %\caption{}
         \label{subfig:astrometryb}}
    \hspace{1em}
    \subfloat[]{ \centering
         \includegraphics[width=0.31\linewidth]{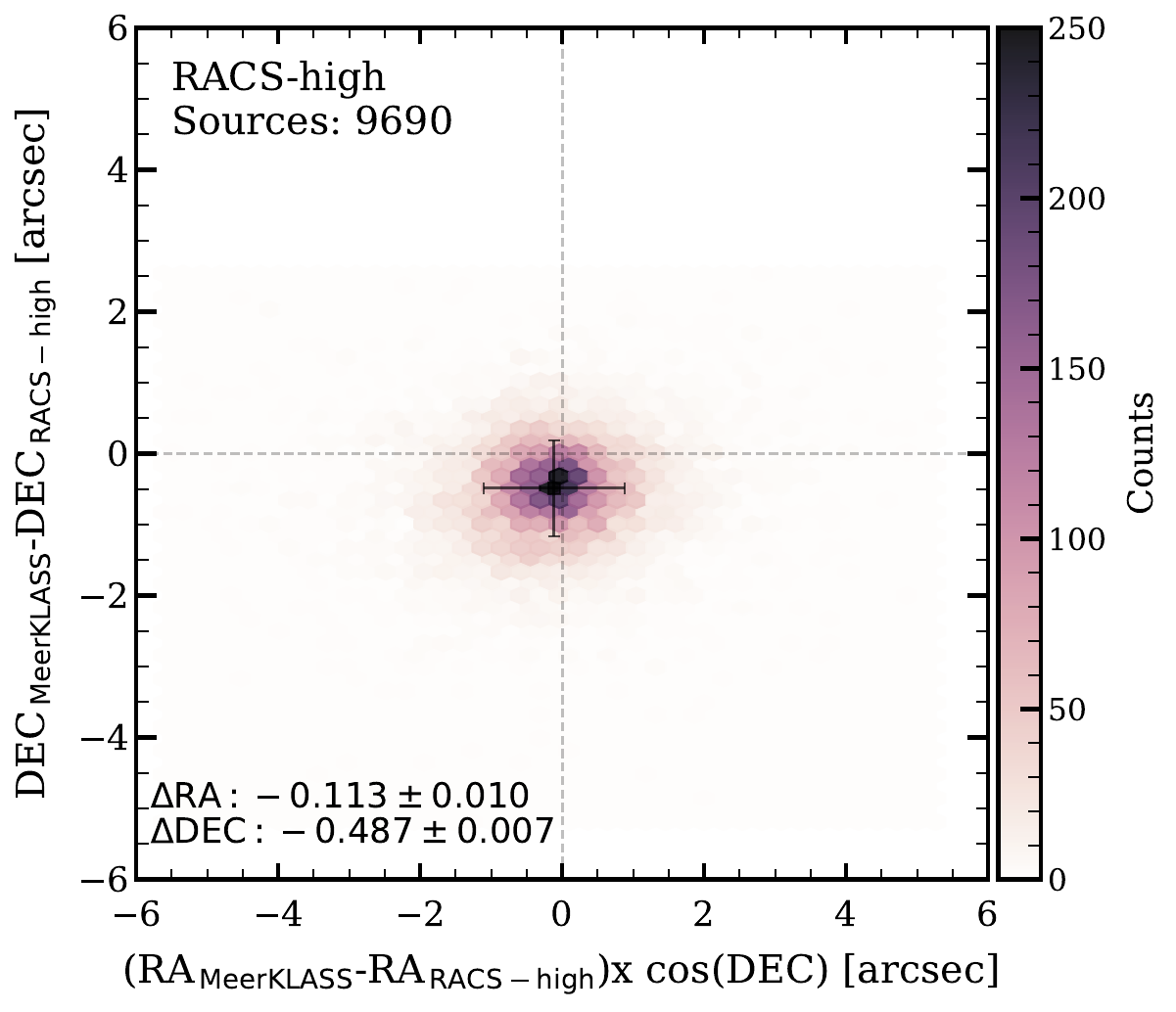}
         %\caption{}
         \label{subfig:astrometry_racshigh}}
    \vspace{-0.15in}
    \subfloat[]{ \centering
         \includegraphics[width=0.3\linewidth]{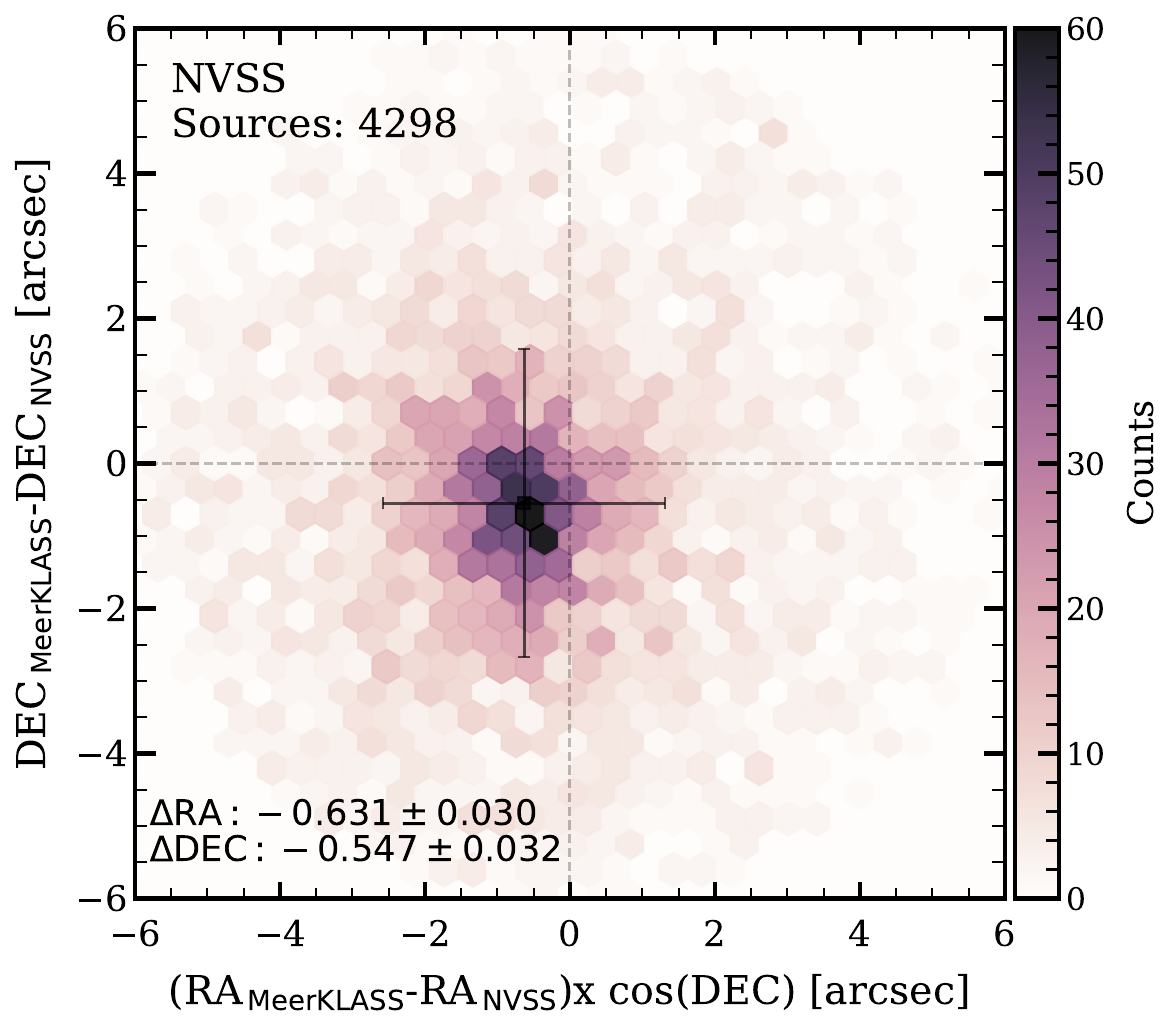}
         %\caption{}
         \label{subfig:astrometryc}}
    \hspace{1em}
    \subfloat[]{ \centering
         \includegraphics[width=0.3\linewidth]{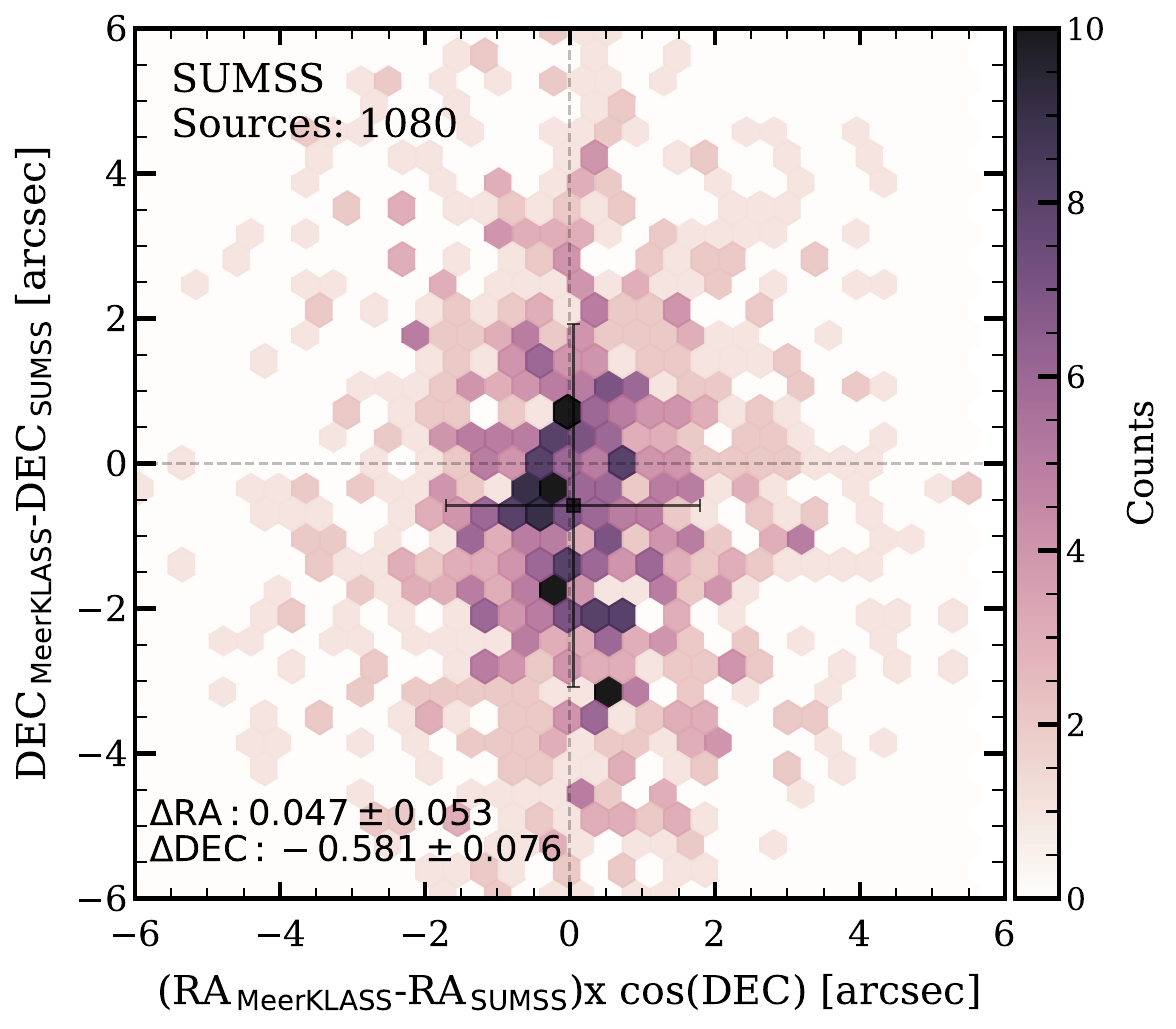}
         %\caption{}
         \label{subfig:astrometryd}}
    \hspace{1em}
    \subfloat[]{ \centering
         \includegraphics[width=0.3\linewidth]{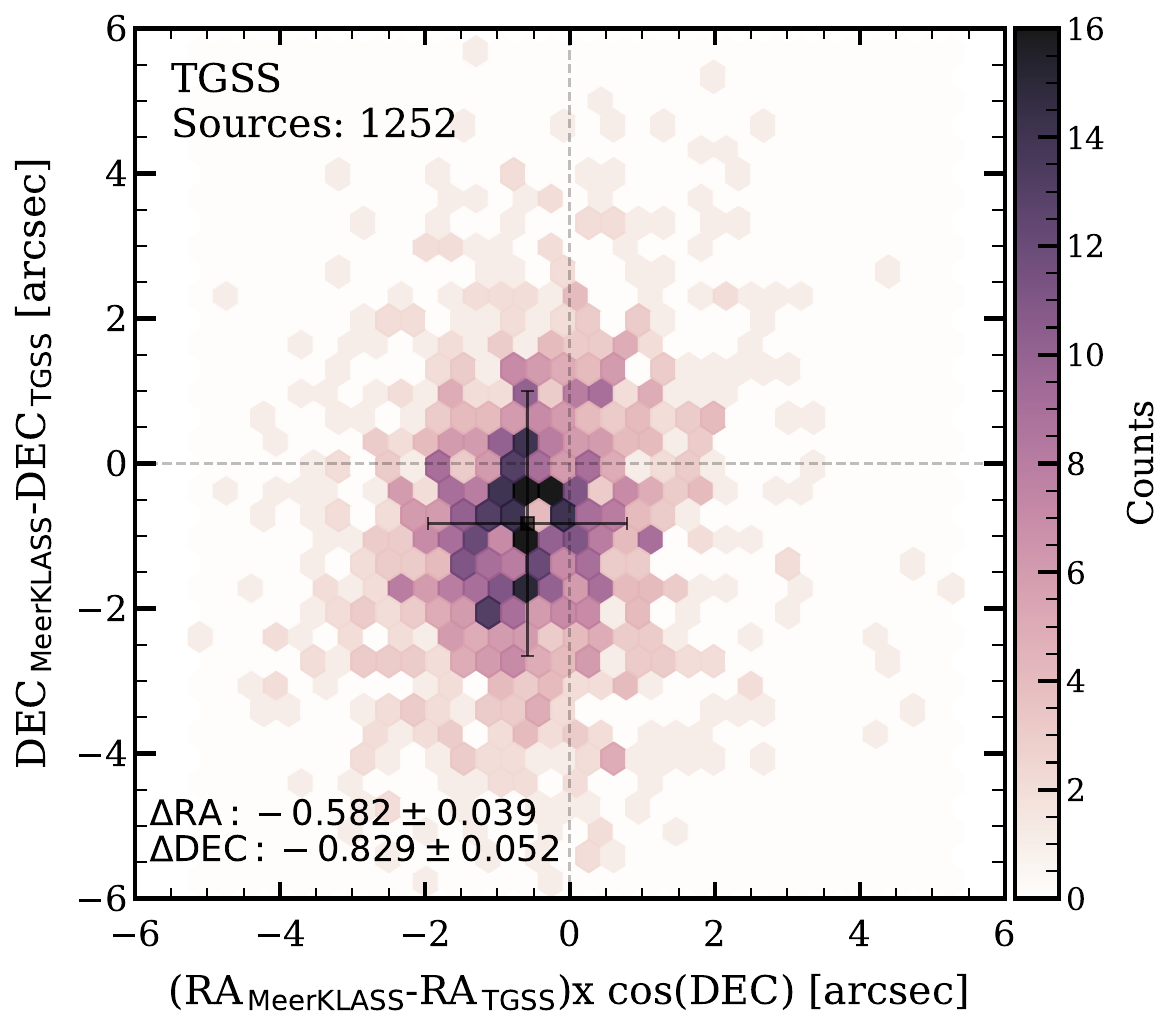}
         %\caption{}
         \label{subfig:astrometry_gleam}}
    \vskip-4pt
    \caption{RA and DEC positional offsets for sources matched between MeerKLASS L-band and the comparison surveys. The colour scale shows the number of matched sources per bin. The RMS scatter of the offsets in each coordinate is denoted by black crosshairs indicating the $1\sigma$ RMS in RA and DEC. Median offsets and their uncertainties are quoted in each panel and summarised in \autoref{tab:comparison_with_cats}. The table also lists the $1\sigma$ RMS scatter of the offsets in (RA, DEC). }
    \label{fig:astrometry}
\end{figure*}

\subsubsection{Identifying unresolved sources}
\label{sec:unresolved}

To decide whether a source is resolved, we follow the standard approach of using the relation between total and peak flux density as a function of SNR \citep[e.g.][]{Bondi2008, LoTSS}. We start from the source catalogue and select only single-component sources (\texttt{S\_code = `S'}) with SNR$\geq 9$, where we define

\[
{\rm SNR} = \frac{S_{\rm peak}}{1.5\,\sigma_{\rm isl}},
\]

i.e. the peak flux density $S_{\rm peak}$ divided by 1.5 times the local island rms noise $\sigma_{\rm isl}$ from \texttt{PyBDSF}. The factor of 1.5 accounts for the effect of OTF scanning on compact sources: time averaging while the array moves broadens the point-source response in right ascension, but does not change the noise pattern on the map. As a result, the flux error of a point source is about 1.5 times larger than the local map RMS. We therefore fold this factor into the SNR definition above and use it consistently for all SNR cuts and flux–error estimates in this work (see Appendix~\ref{app:noise_psf} for details). For these sources we then examine the ratio of integrated flux density $S_{\rm total}$ to peak flux density, $S_{\rm total}/S_{\rm peak}$, as a function of SNR (see \autoref{fig:unresolved}).

For an unresolved source, we expect $S_{\rm total} \approx S_{\rm peak}$, such that $S_{\rm total}/S_{\rm peak} \approx 1$ by definition. In practice, this ratio exhibits some scatter around unity, particularly at low SNR where noise fluctuations affect the measured peak. As can be seen in \autoref{fig:unresolved}, the ratio $S_{\rm total}/S_{\rm peak}$ in our dataset asymptotically approaches a value of 1.06 at high SNR. This small systematic offset likely arises from minor uncorrected effects such as residual gain errors and/or slight astrometric mismatches between overlapping beams. Following established methods adopted in previous studies \citep{Bondi2008,LoTSS}, we define the boundaries of the unresolved-source envelope with an empirical relation of the form 

\begin{equation}
    \left(\frac{S_{\rm total}}{S_{\rm peak}}\right)_{\pm} = 1.06 \pm A\,\times\,{\rm SNR}^{-B},
    \label{eq:envelope_eq}
\end{equation}

and treat sources as resolved when they lie clearly above the upper curve. To measure $A$ and $B$, we first construct the lower envelope.

To build the envelopes, we bin the sample by equally spaced logarithmic SNR and, for each bin with at least 10 sources to keep the percentile estimate stable, we take the $S_{\rm total}/S_{\rm peak}$ value that includes 95\% of objects with $S_{\rm total}/S_{\rm peak}<1.06$, restricting to a narrow band around unity to avoid obvious resolved emission. We then fit these points with \autoref{eq:envelope_eq} using the \texttt{Scipy} function \texttt{curve\_fit} and a non-linear least-squares fit. This yields

\[
    \frac{S_{\rm total}}{S_{\rm peak}} = 1.060 \mp 1.244\,\times\,{\rm SNR}^{-0.845},
\]

where the minus sign defines the lower envelope and the plus sign the corresponding upper envelope. Sources between the two curves are classified as unresolved, and sources above the upper curve are classified as resolved. 

Applying this rule, the blue points in \autoref{fig:unresolved} mark sources classified as unresolved and the grey points mark resolved sources. From this we estimate that approximately 53.5\% of MeerKLASS sources are unresolved, and therefore they should also appear unresolved in the comparison catalogues at similar or coarser resolution.

\begin{figure*}%[!h]
     \centering
     \subfloat[]{
         \includegraphics[width=0.31\linewidth]{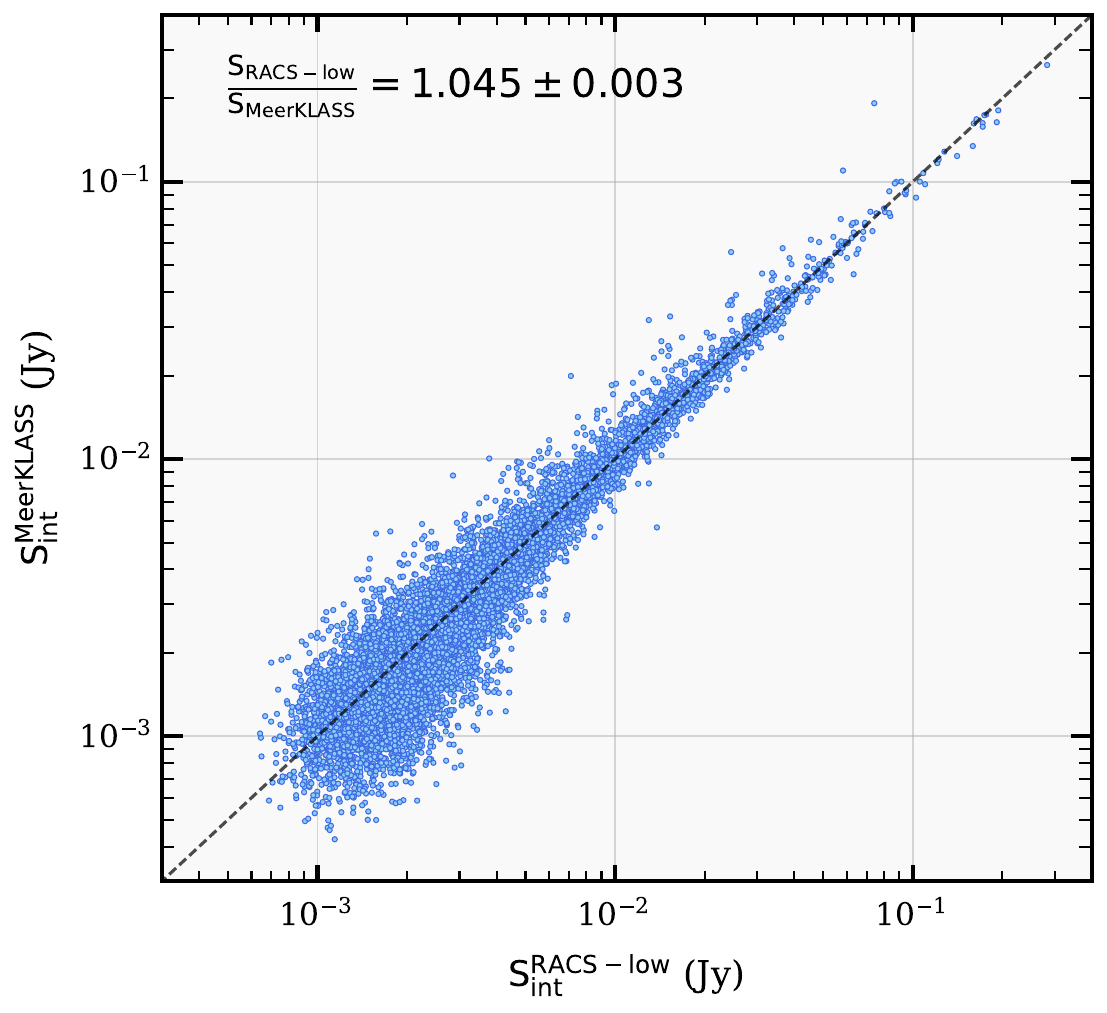}
         %\caption{hi}
         \label{subfig:intfluxa}}
    \hspace{1em}
    \subfloat[]{ \centering
         \includegraphics[width=0.31\linewidth]{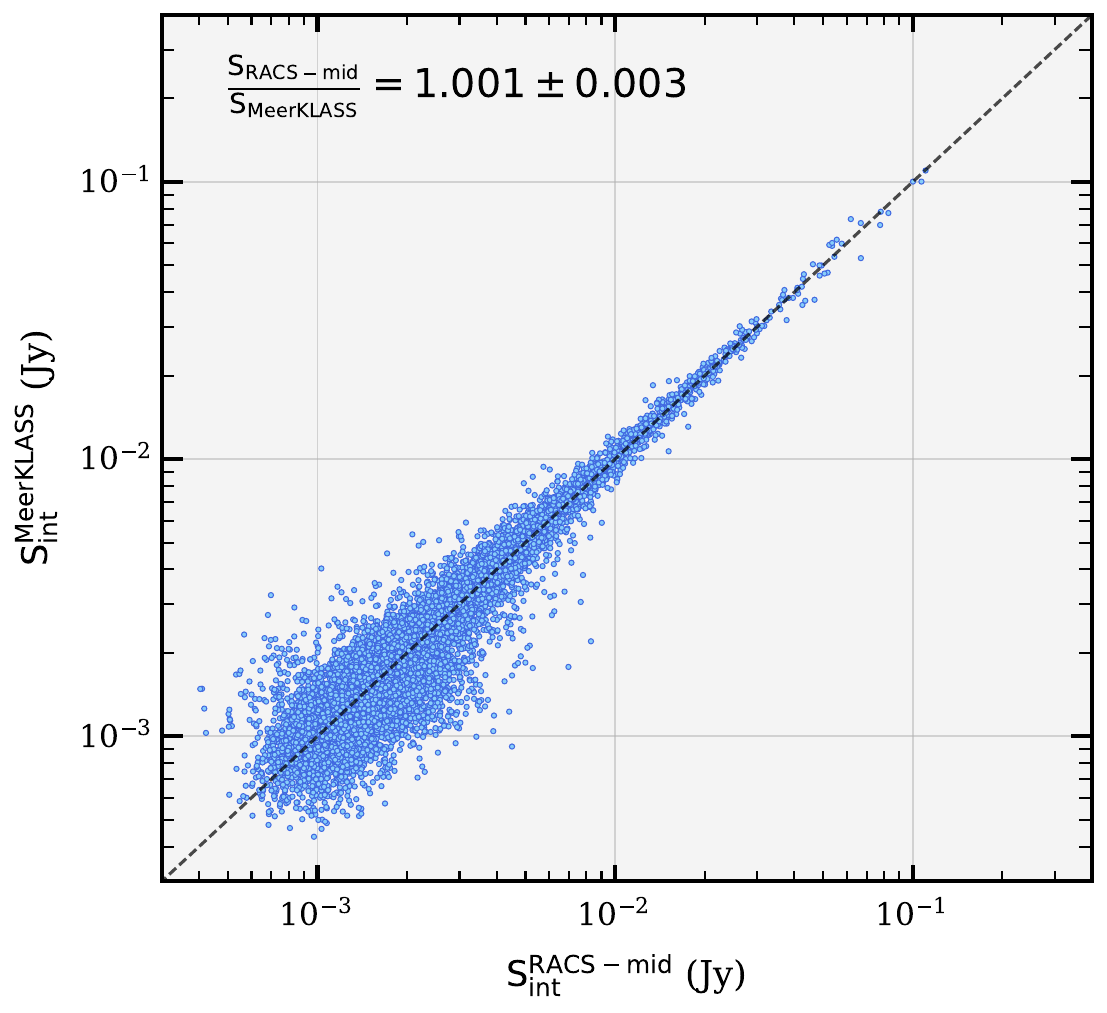}
         %\caption{}
         \label{subfig:intfluxb}}
    \hspace{1em}
    \subfloat[]{ \centering
         \includegraphics[width=0.31\linewidth]{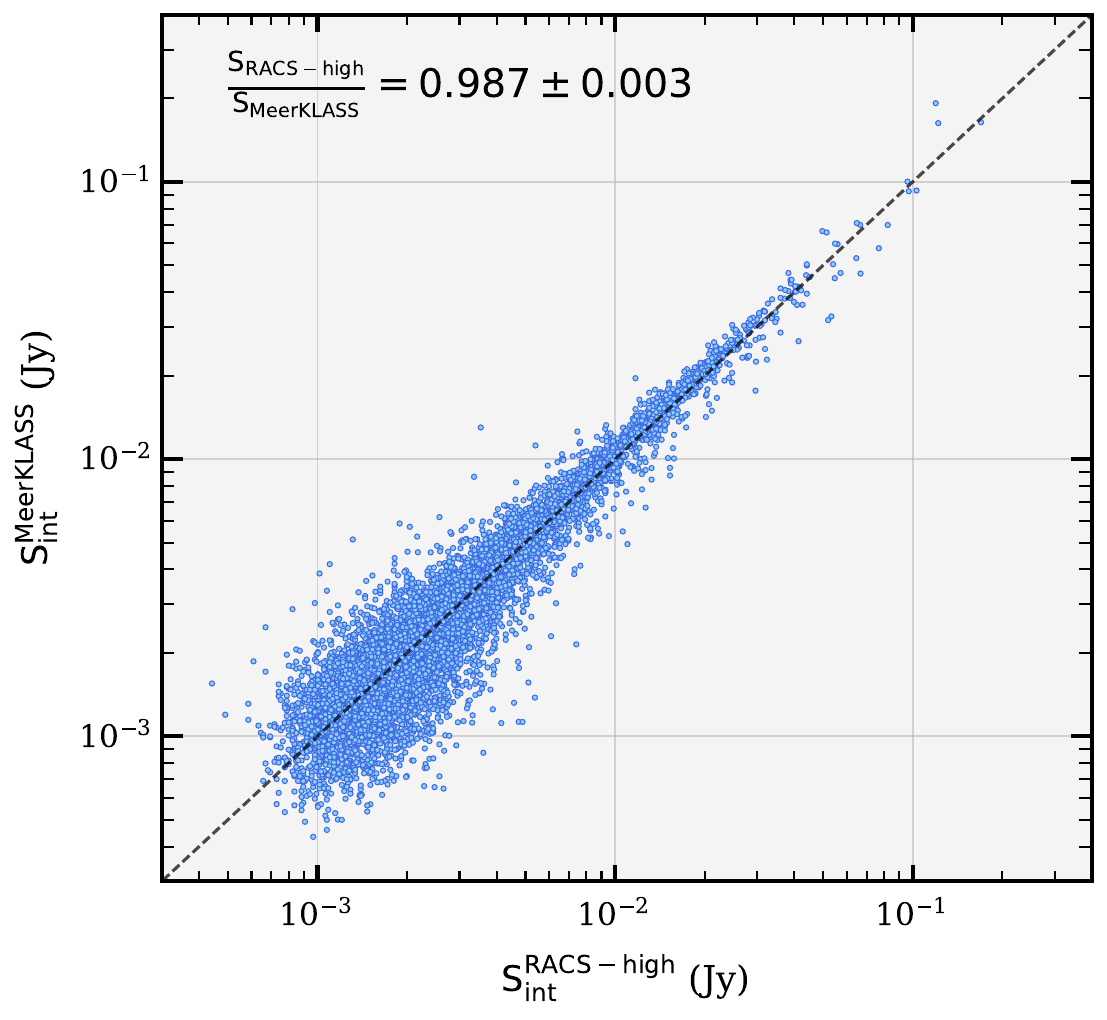}
         %\caption{}
         \label{subfig:intfluxc}} 
    \vspace{-0.1in}
    \subfloat[]{ \centering
         \includegraphics[width=0.31\linewidth]{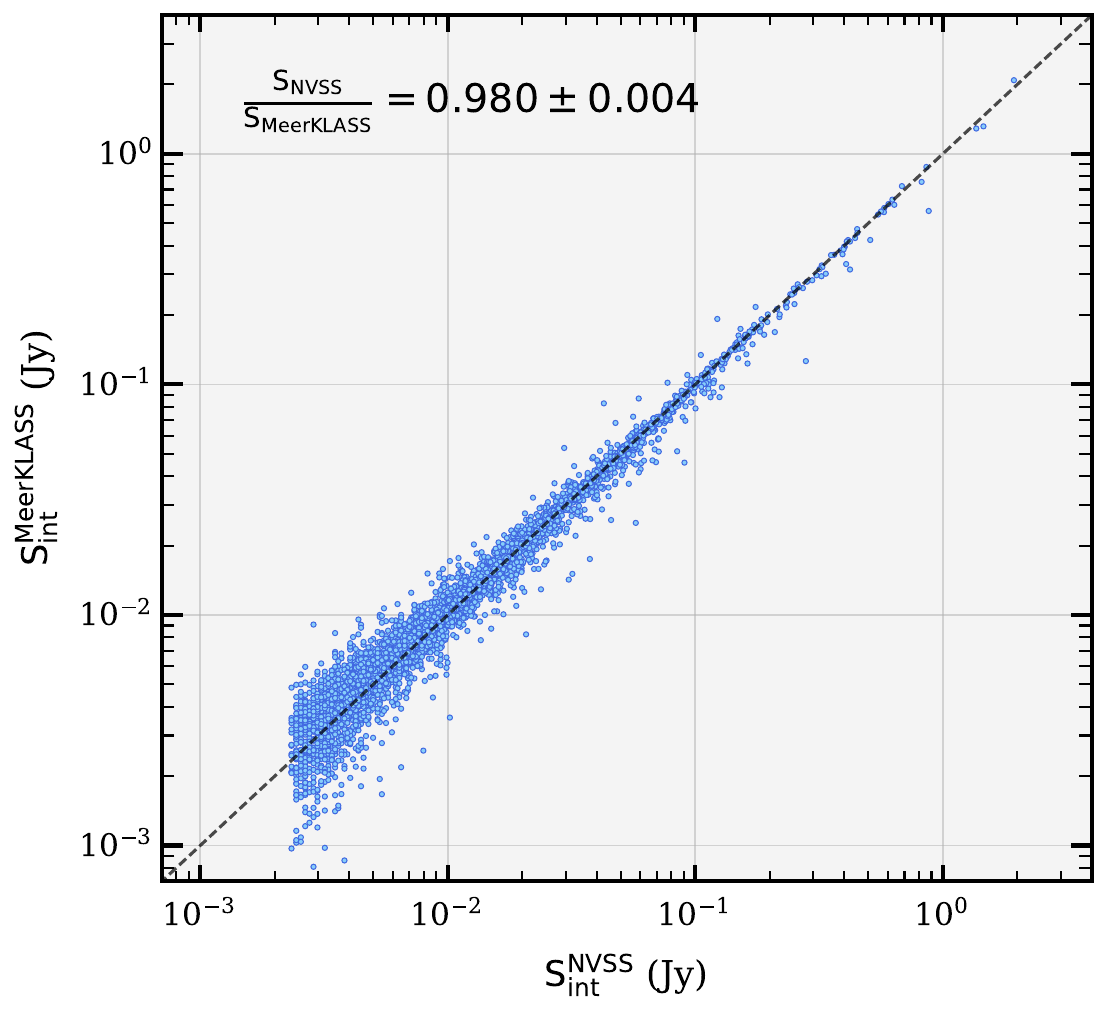}
         %\caption{}
         \label{subfig:intfluxd}}
    \hspace{1em}
    \subfloat[]{ \centering
         \includegraphics[width=0.31\linewidth]{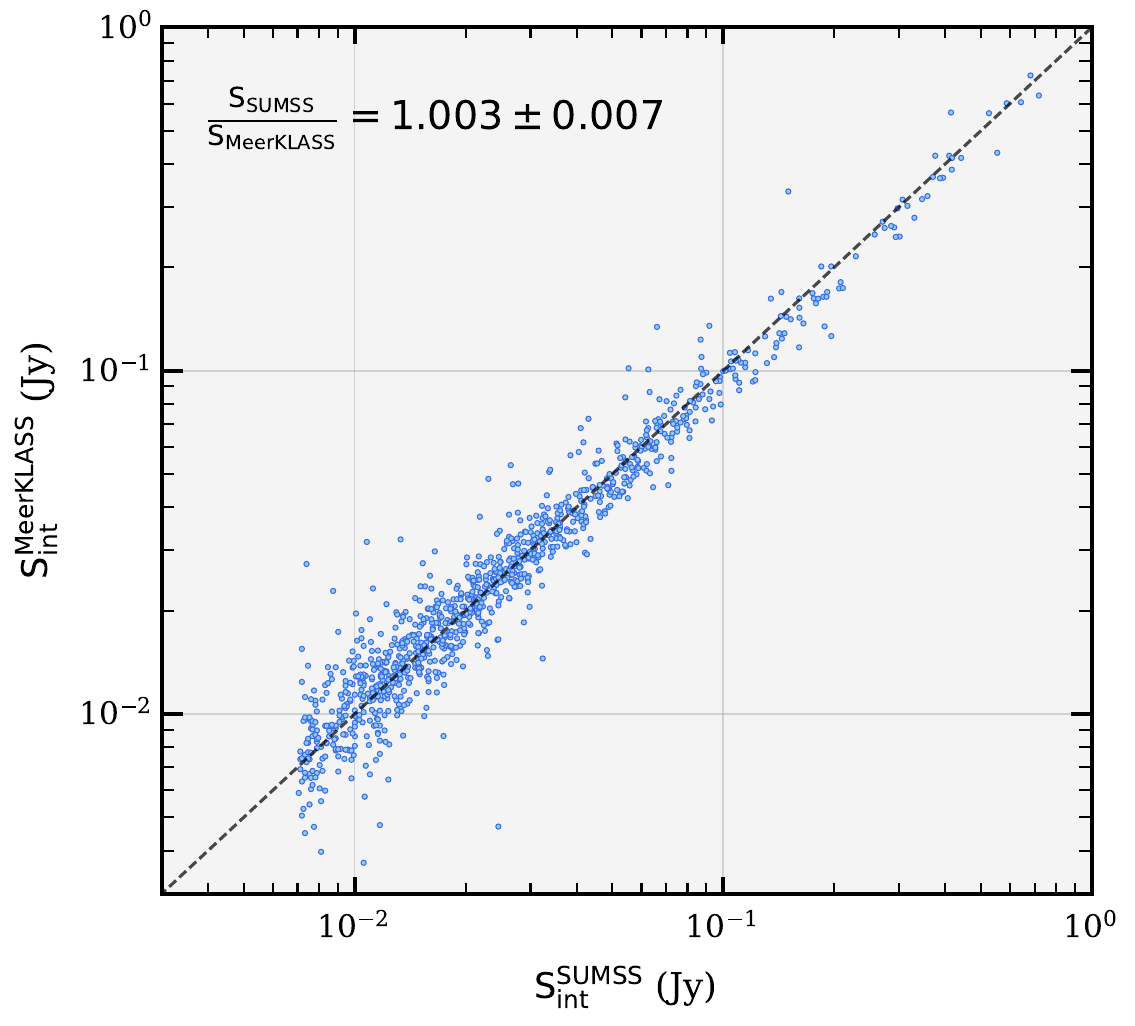}
         %\caption{}
         \label{subfig:intfluxe}}
    \hspace{1em}
    \subfloat[]{ \centering
         \includegraphics[width=0.31\linewidth]{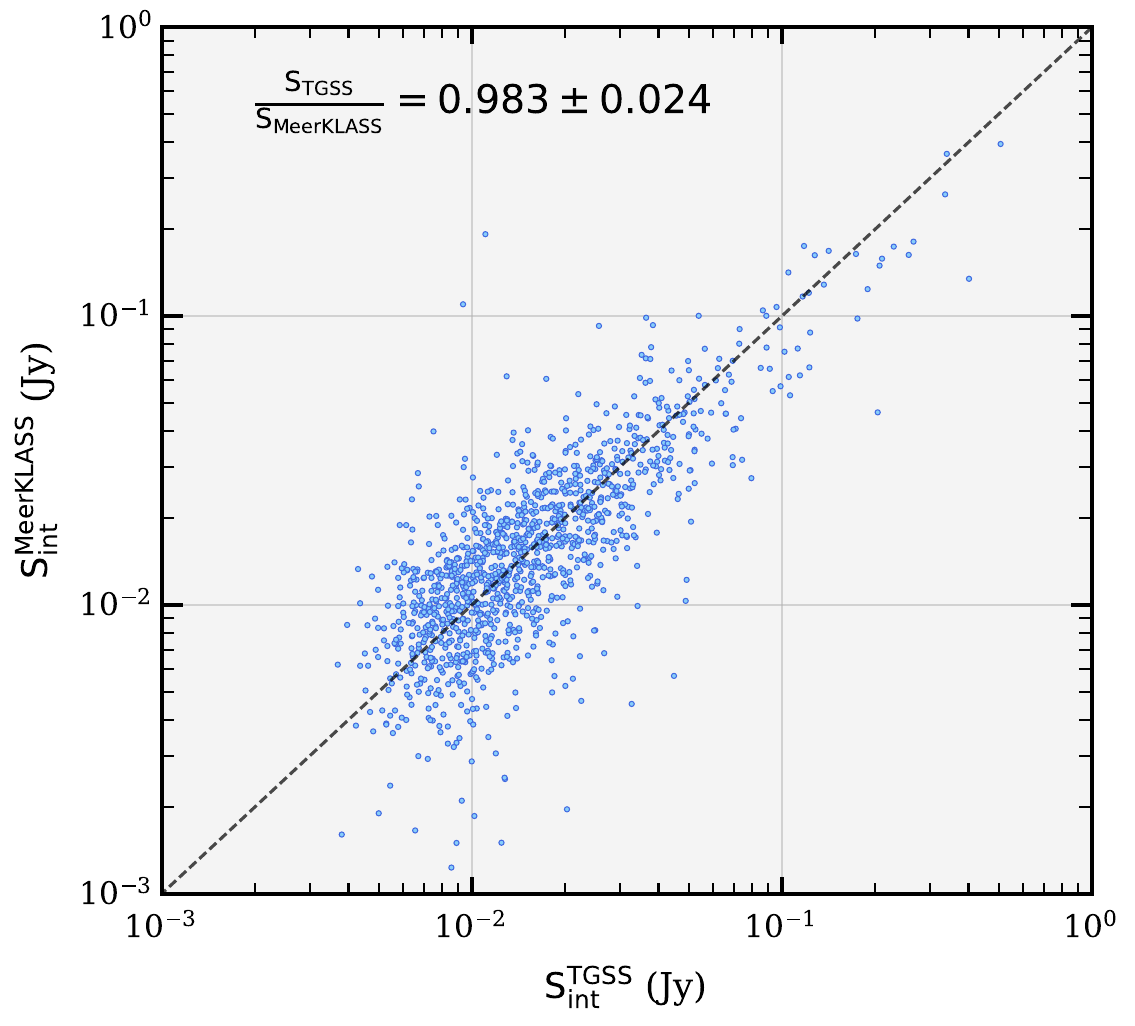}
         %\caption{}
         \label{subfig:intfluxf}}
    \caption{Comparison of the ratio of fluxes from other surveys to our measured fluxes for ensembles of the matched sources (see section \ref{sec:matching_cats}). The flux ratios are listed in each sub-figure and in \autoref{tab:comparison_with_cats}; the ratios are close to unity for all surveys.}
    \label{fig:compare_flux}
\end{figure*}

\begin{table*}
    \centering
    \caption{Astrometry and flux-density comparisons of the MeerKLASS L-band survey (1284 MHz) with other surveys. $\sigma_{\rm RA}$ and $\sigma_{\rm DEC}$ are the $1\sigma$ RMS scatter of the positional offsets in RA and DEC, respectively. The flux ratio reports the median $S_{\rm survey}/S_{\rm MeerKLASS}$ (scaled to 1284\,MHz). Uncertainties are estimated as the $\sigma/\sqrt{N_{\rm sources}}$.}
    
    \begin{tabular}{lrrrcrcc}
        \hline
        Survey & \multicolumn{1}{c}{Frequency} & \multicolumn{1}{c}{N$_{\rm sources}$} & \multicolumn{1}{c}{RA Offset} & \multicolumn{1}{c}{$\sigma_{\mathrm{RA}}$} & \multicolumn{1}{c}{DEC Offset}  & \multicolumn{1}{c}{$\sigma_{\mathrm{DEC}}$} & Flux ratio \\
        & \multicolumn{1}{c}{(MHz)}     &                                       & \multicolumn{1}{c}{(arcsec)}  & \multicolumn{1}{c}{(arcsec)}         & \multicolumn{1}{c}{(arcsec)}   & \multicolumn{1}{c}{(arcsec)}                       & \\
        \hline
        RACS-low & 887.5 & 11,991 & $-0.137\pm0.014$  &  1.500 & $-0.872\pm 0.013$ & 1.430 & $1.045 \pm 0.003$\\ 
        RACS-mid & 1367.5 & 11,243 & $ -0.015 \pm 0.010$ &  1.027 & $-0.121 \pm 0.007$ & 0.693 & $1.001 \pm 0.003$ \\ 
        RACS-high & 1655.5 & 9690 & $-0.113 \pm 0.010$ & 0.994 & $-0.487 \pm 0.007$ & 0.674 & $0.988 \pm 0.004$ \\ 
        NVSS & 1400.0 & 4,298 & $-0.631 \pm 0.030$ & 1.946 & $-0.547 \pm 0.032$ & 2.128 & $ 0.980 \pm 0.004$ \\
        SUMSS  & 843.0 & 1,080 & $ 0.047 \pm 0.053$ & 1.751 & $-0.581 \pm 0.076$ & 2.499 & $1.003 \pm 0.007$ \\
        TGSS-ADR & 150.0 & 1,252 & $-0.582 \pm 0.039$ & 1.375 & $-0.829 \pm 0.052$ & 1.824 & $0.983 \pm 0.024$\\   
        \hline
    \end{tabular}
    \label{tab:comparison_with_cats}
\end{table*}

\subsubsection{Catalogue matching}
\label{sec:matching_cats}

For a reliable comparison, we apply a set of selection criteria to minimize systematic biases arising from complex source morphology, different angular resolutions, and source blending. Specifically, our applied criteria are:

\begin{enumerate}
    \item Morphology: We select only sources classified as single-Gaussian components in PyBDSF. In the MeerKLASS catalogue, this corresponds to objects with \texttt{S\_code = `S'}, indicating that the source is modelled by a single Gaussian. A corresponding classification is applied to the comparison catalogue aswell.
    \item Source isolation: Sources are required to be isolated within an angular separation equal to twice the larger of the two catalogue's FWHM values. The same isolation criterion is applied to the comparison survey. Accordingly, when comparing with RACS-low, -mid, -high, and TGSS, the isolation radius is $50\arcsec$, whereas for NVSS and SUMSS the corresponding isolation radius is $90\arcsec$.
    \item SNR: Robust detections with SNR\,$\geq$\,9 in the MeerKLASS catalogue.
    \item Match radius: Cross-matching between catalogues is performed within a $6\arcsec$ radius, corresponding to approximately four MeerKLASS L-band image pixels. This accounts for minor positional uncertainties while keeping spurious associations to a minimum.
\end{enumerate}

Sources meeting these conditions are used to evaluate positional and flux-density offsets, as well as to derive spectral indices. The spectral index, $\alpha$, characterises the broadband radio emission as a power law, $S_{\nu} \propto \nu^{\alpha}$, where $S_{\nu}$ is the flux density at frequency $\nu$. In the synchrotron-dominated regime, typical values of $\alpha$ lie between $-0.7$ and $-0.8$.

\subsubsection{Astrometric offsets}

For each source with a matched counterpart in an external survey we compute the Right Ascension and Declination offsets as the difference between the MeerKLASS position and the external catalogue position. The resulting offset distributions are tightly clustered around the origin, with RMS scatters of $\sim 1\arcsec-1.5\arcsec$ for the three RACS bands and $\sim 1.5\arcsec-2.5 \arcsec$ for NVSS, SUMSS and TGSS, indicating good overall positional agreement. The median offsets, however, are small but statistically significant in all cases because the uncertainties on the median are at the $\sim 0.01\arcsec-0.05\arcsec$ level, allowing us to resolve even very small systematic differences in astrometry.

Among the RACS surveys, our astrometry aligns best with RACS-mid: the mean declination offset is only $-0.121\pm 0.007~\arcsec$ and the RA offset is consistent with zero within the quoted uncertainty ($-0.015 \pm 0.010~\arcsec$). RACS-low and RACS-high show larger systematic shifts in declination, with mean offsets of $-0.872 \pm 0.013~\arcsec$ and $-0.487 \pm 0.007~\arcsec$, respectively, and RA offsets of order $-0.1\arcsec$. This implies internal astrometric differences at the few-tenths of an arcsecond level between the RACS bands themselves. Relative to NVSS we find coherent offsets of $\simeq -0.6\arcsec$ in both RA and Dec, while SUMSS and TGSS show similar sub-arcsecond shifts. The mean offsets for all surveys are summarised in \autoref{tab:comparison_with_cats}.

Despite these small but significant systematic differences with individual reference catalogues, the absolute MeerKLASS L-band DR1 astrometry remains accurate at the sub-arcsecond level: all median offsets are $\lesssim 1\arcsec$ in both coordinates. This is comparable to other MeerKAT L-band products such as MALS-DR1, where typical offsets are of order $1\arcsec$ \citep[see][]{MALs_Deka_2024}. This level of positional accuracy is sufficient for robust multi-wavelength cross-matching with optical and infrared surveys.

\begin{figure}
    \centering
    \includegraphics[width=1\linewidth]{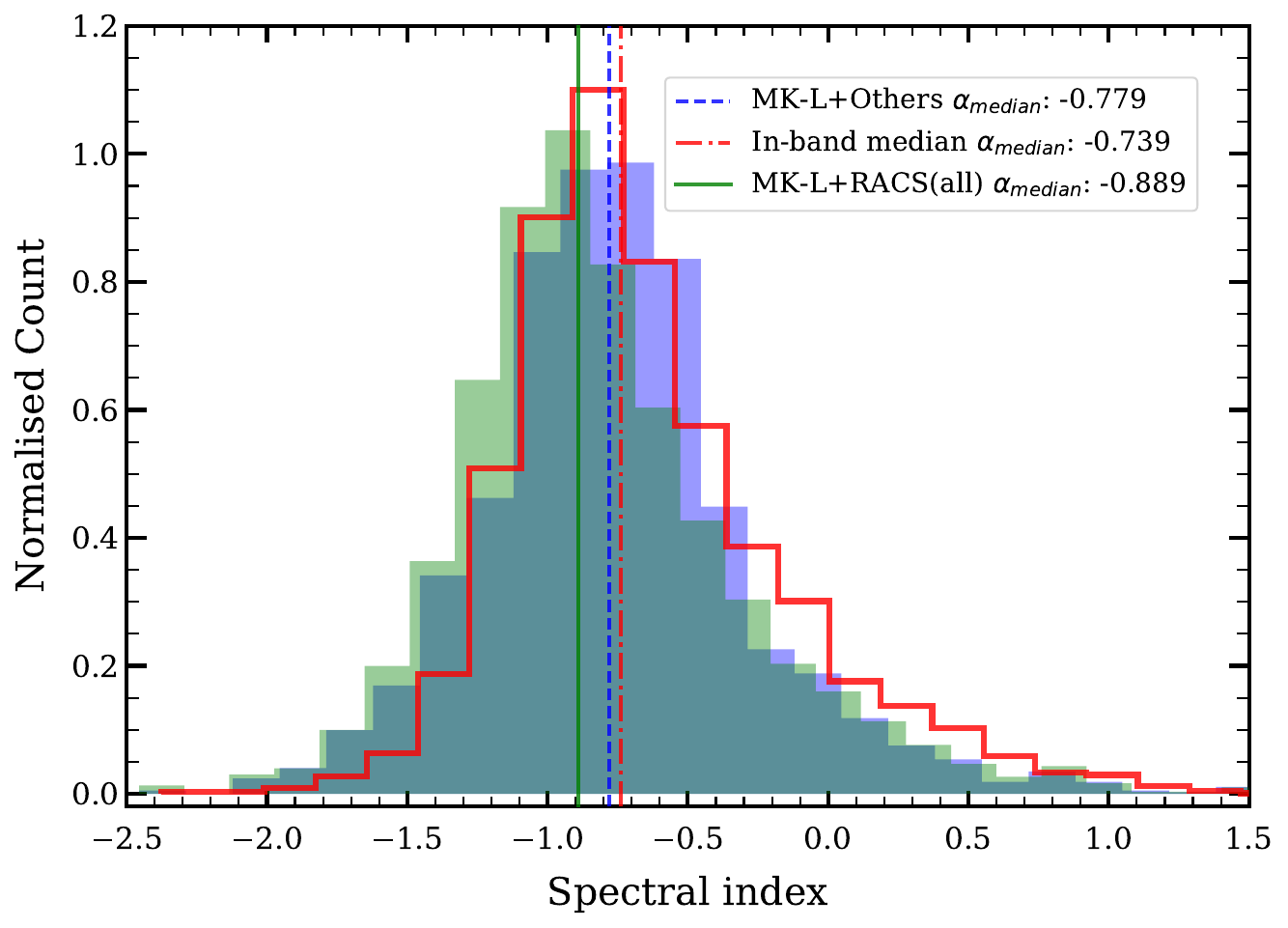}
    \vskip-6pt
    \caption{Spectral index distribution measured for MeerKLASS L-band (inband; red), MeerKLASS L-band with RACS-low -mid and -high (green), and MeerKLASS L-band plus additional surveys (see text). To ensure reliable spectral index measurements, a sub-sample of sources are selected by imposing a constraint on the spectral index uncertainty  $\sigma_\alpha < 0.2$. }
    \label{fig:SED_histogram}
\end{figure}

\begin{figure*}
    \centering
    \includegraphics[width=0.31\linewidth]{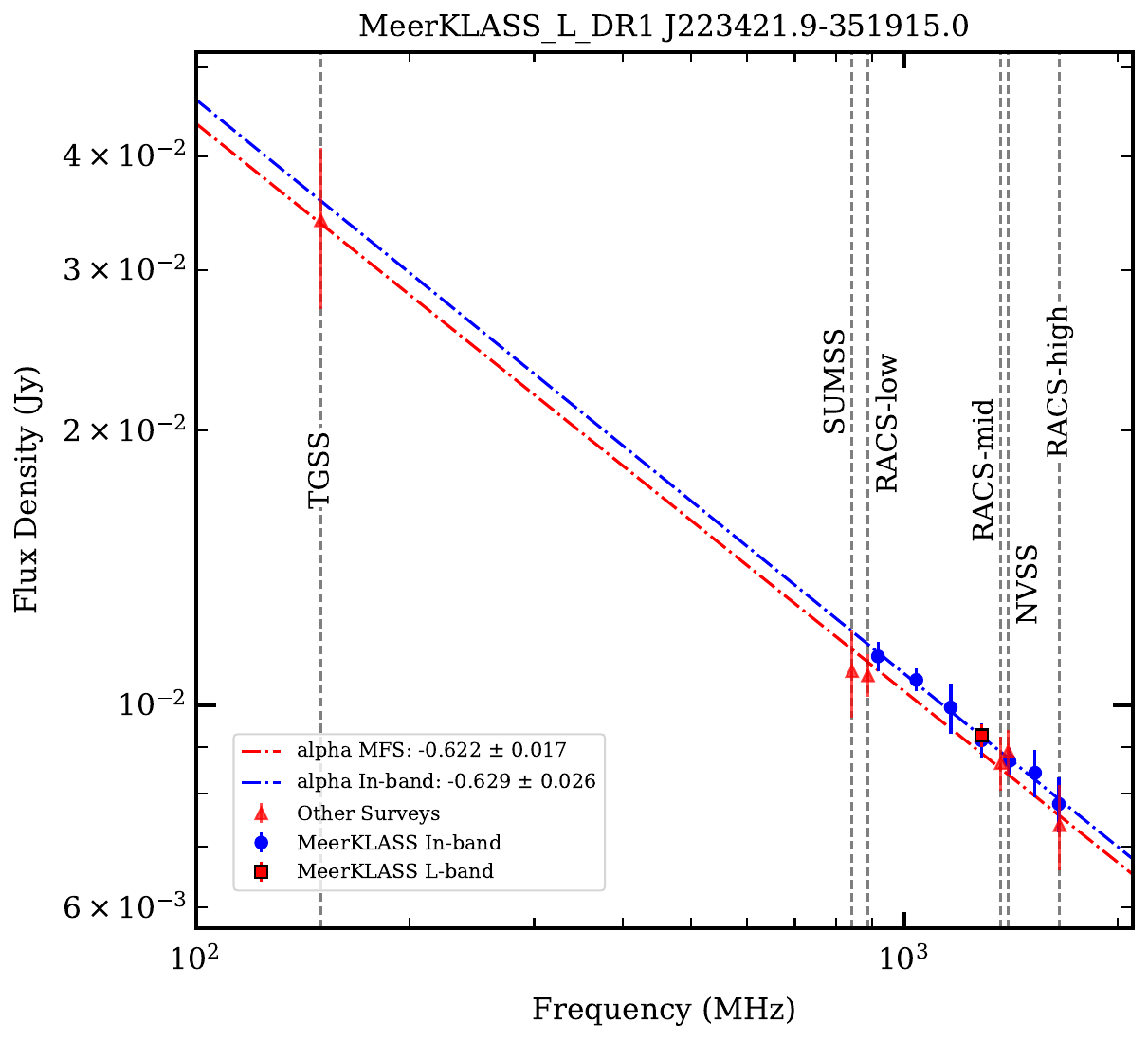}
    \hspace{1em}
    \includegraphics[width=0.31\linewidth]{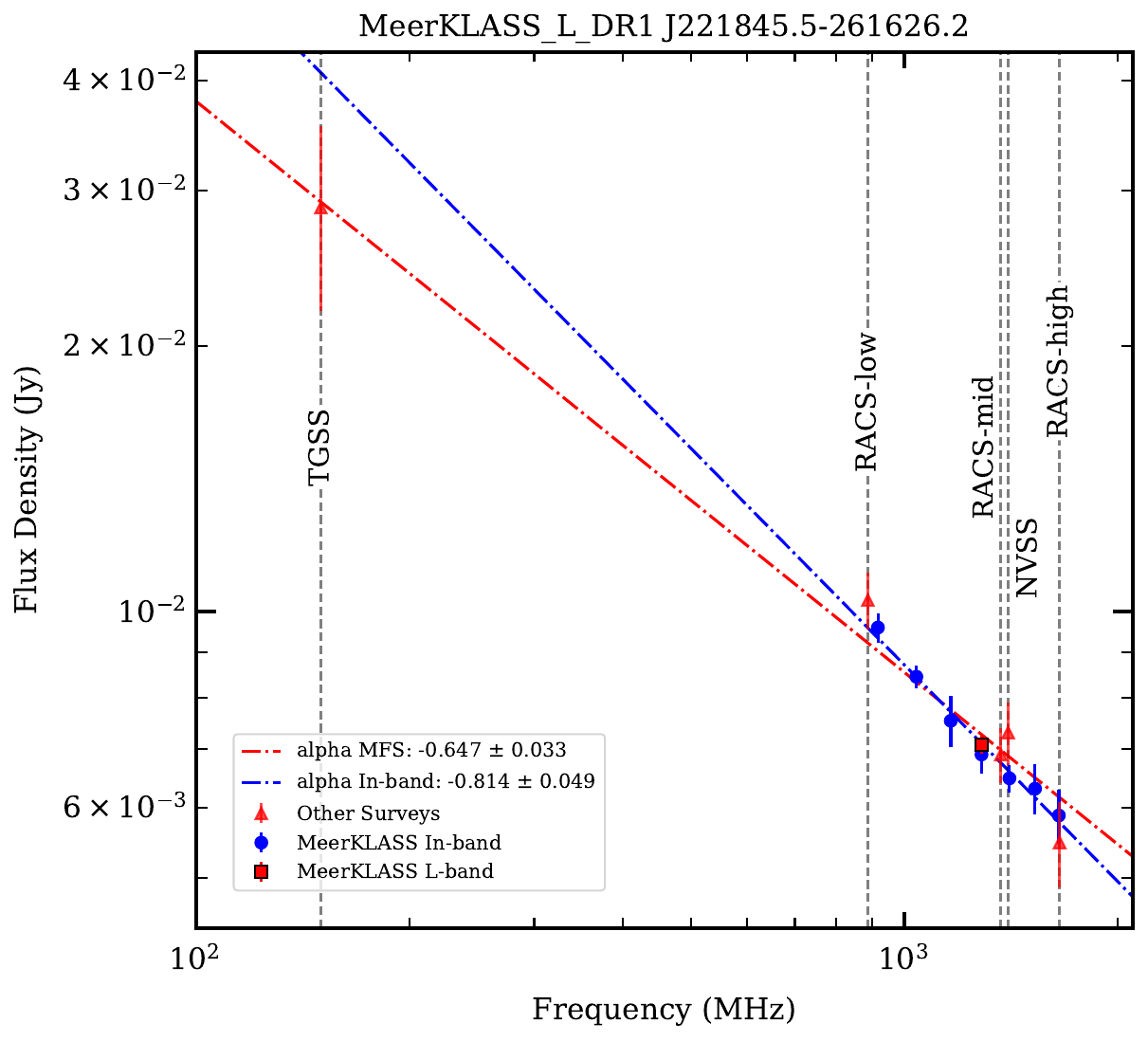}
    \hspace{1em}
    \includegraphics[width=0.31\linewidth]{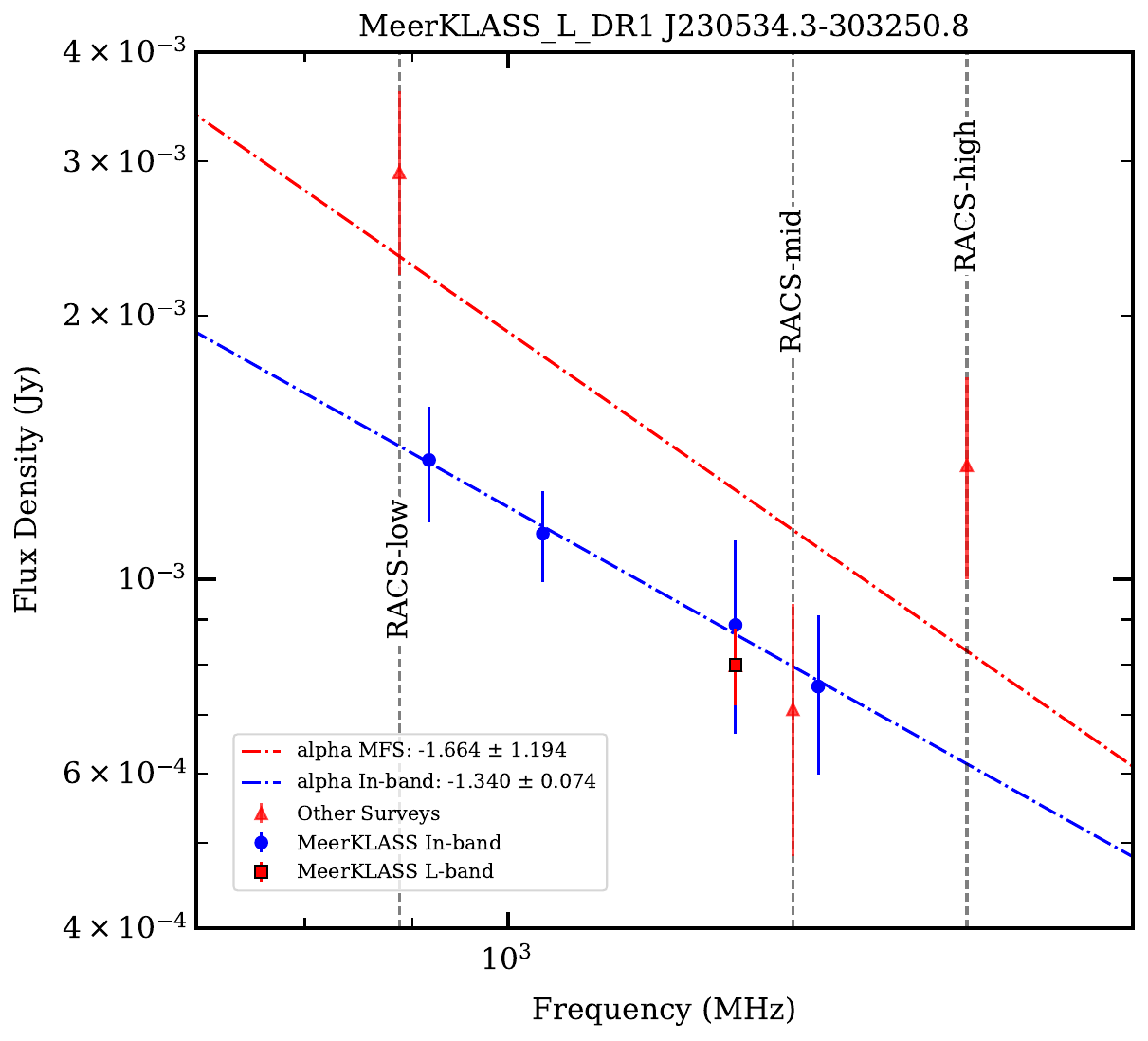}
    \vskip-6pt
    \caption{SEDs for three representative MeerKLASS L-band sources cross-matched with external radio surveys (vertical dashed lines).  Flux densities from other surveys are broadly consistent with these fits but have noticeably larger fractional errors, especially for the faintest $\sim0.8$\,mJy source in the right-hand panel, where several external measurements are close to their survey limits. The MeerKLASS sub-band measurements provide several in-band SED samples across the wide L-band bandwidth, enabling the spectral index to be derived from MeerKLASS data alone and minimising the impact of cross-survey calibration offsets and source variability.}
    \label{fig:sed_example1}
\end{figure*}

\subsubsection{Flux offsets}

We compared the integrated flux densities ($S_{\rm int}$) using the same set of matched sources. To enable direct comparison, the flux densities from each external survey are scaled to the effective mean MeerKLASS L-band frequency of 1284\,MHz using a power-law relationship ($S_{\nu} \propto \nu^{\alpha}$). Rather than adopting a single spectral index for all catalogues, we use the spectral indices assumed in the original survey papers wherever possible, so that we remain consistent with the flux scales established by the comparison catalogues themselves. For RACS-low (887.5\,MHz) and SUMSS (843\,MHz) we adopt $\alpha=-0.9$ and $\alpha=-0.83$, respectively, while for RACS-mid, RACS-high, NVSS and TGSS we use $\alpha=-0.7$, appropriate for typical steep-spectrum radio sources.

Flux density comparisons against RACS-mid are of particular interest, because its observing frequency is close to the MeerKLASS L-band. We adopt $S\,\propto\,\nu^{\alpha}$ with a nominal $\alpha=-0.7\pm0.1$, so the RACS-mid to MeerKLASS frequency shift implies a flux change of only $\pm0.5\%$, i.e., well below the statistical precision of our flux-ratio measurements. For NVSS at 1.4\,GHz, the same $\pm0.1$ error in $\alpha$ leads to about $\sim\pm5\%$ uncertainty in the flux estimates at the MeerKLASS mean frequency, which is comparable to the absolute flux-scale uncertainty of that survey.

In each panel of \autoref{fig:compare_flux} we plot $S_{\rm int}$ from the external survey against the corresponding MeerKLASS $S_{\rm int}$ on log–log axes, with the dashed line showing the 1:1 relation. We quantify the photometric agreement using the flux-density ratio $R = S_{\rm ext}/S_{\rm MeerKLASS}$ for each matched source, and we report the median ratio and its uncertainty (computed as ${\rm NMAD}/\sqrt{N_{\rm sources}}$) in both the figure and \autoref{tab:comparison_with_cats}. 
The same table also lists the survey frequencies and the number of matched sources contributing to each comparison.

All six surveys yield mean flux-density ratios within 5\% of unity. RACS-mid is essentially on the MeerKLASS scale, with $R = 1.001 \pm 0.003$. RACS-low and RACS-high show small but highly significant offsets, with ratios of $1.045 \pm 0.003$ and $0.988 \pm 0.004$, respectively, indicating systematic differences at the $\sim 2$–5\% level. The NVSS ratio of $0.980 \pm 0.004$ similarly reflects a $\sim 2\%$ offset. SUMSS ($1.003 \pm 0.007$) and TGSS ($0.983 \pm 0.024$) are consistent with unity within their larger uncertainties, with TGSS also showing the largest intrinsic scatter, as expected for the lowest-frequency comparison.

Taken together, these results demonstrate an accuracy in the photometric scale of better than $\mathbf{5\%}$ over the $\sim 0.15$–1.7\,GHz frequency range probed by the comparison surveys, which is within the $\sim 10\%$ absolute flux density uncertainty typical for GHz surveys \citep{Duchesne_RACS_MID_DR2}. The tight scatter in the flux to flux comparisons in \autoref{fig:compare_flux} and the resulting small uncertainty in the median ratio presented there and in \autoref{tab:comparison_with_cats} indicate there are small but real systematic flux offsets. These small systematic flux offsets could be attributed to imperfections in MeerKLASS, in the comparison survey or both.  Overall, these comparisons affirm the stability and fidelity of the MeerKLASS L-band calibration process.  As mentioned previously, ongoing improvements to the OTF imaging pipeline are underway and we expect the photometry to improve further in DR2.
 
\subsection{Spectral index comparisons}
\label{sec:specindex}

To characterize the spectral energy distribution (SED) of the MeerKLASS L-band radio source population, we calculate the spectral index, $\alpha$, for sources that are successfully cross-matched with external surveys at different frequencies and/or detected in multiple MeerKLASS sub-bands. The spectral index is defined by the power law $S_{\nu} \propto \nu^{\alpha}$, where $S_{\nu}$ is the integrated flux density at frequency $\nu$. In the synchrotron-dominated regime (which covers the frequencies used here), extragalactic sources typically exhibit steep spectra with $\alpha$ between $-0.7$ and $-0.8$.

The spectral index, $\alpha$, for each source is obtained by fitting a single power-law model

\begin{equation}
    S_{\nu} = A \left(\frac{\nu}{\nu_0}\right)^{\alpha}
\end{equation}

to its measured flux densities across frequency. In practice we work in log space, fitting

\begin{equation}
    \log_{10} S_{\nu,i} = \log_{10} A + \alpha \,\log_{10}\,\left(\frac{\nu_i}{\nu_0}\right)
\end{equation}

with a weighted least-squares regression over all available measurements $S_{\nu,i}$ at frequencies $\nu_i$. We require at least four independent flux-density points to enter the fit. We perform spectral-index comparisons using measurements from TGSS-ADR1 (150\,MHz), SUMSS (843\,MHz), RACS-low (887.5\,MHz), -mid (1367.5\,MHz), -high (1655.5\,MHz), and NVSS (1400\,MHz), combined with the MeerKLASS L-band integrated and/or in-band sub-band flux densities as described above.

We specifically used sources that met our mentioned matching criteria, focusing on unresolved, isolated, high-SNR sources to mitigate biases from source resolution and noise effects. For all SED fits shown in this section (both in the histograms and the example SEDs), we additionally require that each source has at least four independent flux-density measurements contributing to the fit (e.g., MeerKLASS plus $\ge 3$ external points).

Moreover, because we have imaged in seven subbands in MeerKLASS L-band, we can also fit an in-band spectral index using MeerKLASS alone. For each source with reliable sub-band flux densities, we fit a simple power law to the L-band sub-band points and refer to the resulting slope as the ``in-band'' spectral index, $\alpha_{\rm in-band}$. This avoids relying on external catalogues and therefore removes sensitivity to cross-survey flux-scale differences and long-term source variability.

\autoref{fig:SED_histogram} shows the distribution of spectral indices for three choices of data. The blue histogram (MK-L+Others) combines MeerKLASS with all available external surveys; the green histogram (MK-L+RACS) uses only RACS-low, -mid and -high; and the red histogram shows the in-band MeerKLASS measurements. To ensure robust spectral indices we restrict this comparison to sources with formal spectral-index uncertainties $\sigma_{\alpha} \le 0.2$ and with at least four flux-density measurements in the SED fit. Under these cuts the sample sizes are 2,230 sources for MK-L+Others, 1,865 for MK-L+RACS(all), and 6,237 for the in-band sample. All three distributions peak near $\alpha \simeq -0.8$, consistent with a synchrotron-dominated extragalactic population. The median values are $\alpha_{\rm med} = -0.779$ for MK-L+Others, $-0.739$ for the in-band indices, and $-0.889$ for MK-L+RACS(all). The key result is that the in-band distribution is only slightly steeper than the cross-survey cases, demonstrating that MeerKLASS alone already provides competitive spectral-index measurements for a large fraction of the source population.  

\begin{figure}
    \centering
    \includegraphics[width=1\linewidth]{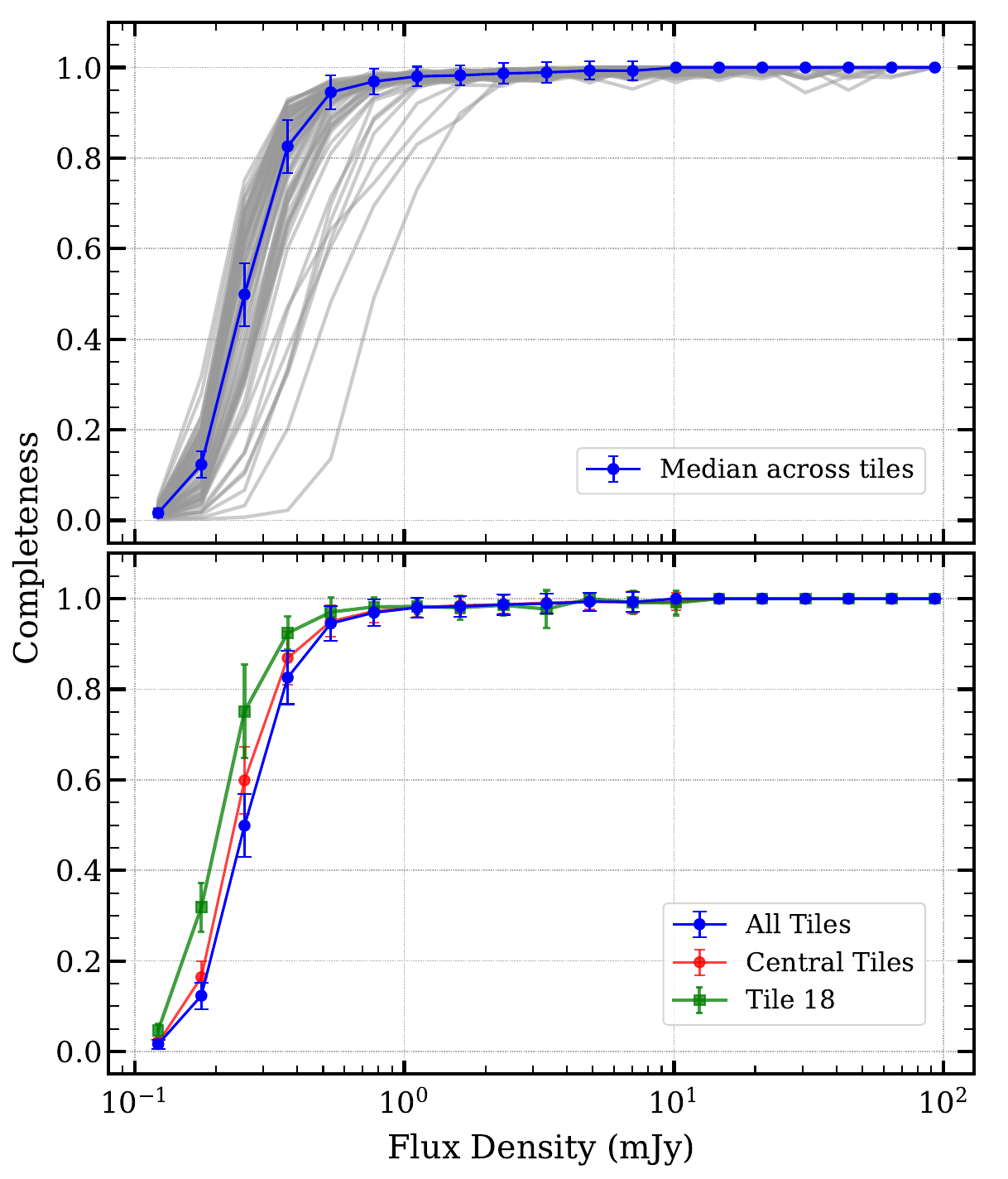}
    \vskip-8pt
    \caption{\textit{Top}: Completeness curve for each tile (grey lines), along with the median curve (with 1 sigma spread) in blue across all 67 tiles. \textit{Bottom}: Completeness curve for the full DR1 survey (blue), central tiles in red, and a tile 18 in green, along with the error bars showing $1\sigma$ across realizations.}
    \label{fig:completness}
\end{figure}

\autoref{fig:sed_example1} illustrates the SEDs of three representative sources. For each source we plot the MeerKLASS L-band integrated flux density and the seven in-band sub-band measurements, along with flux densities from the external surveys at their respective frequencies. Simple power-law fits to the MeerKLASS MFS flux (dashed red line) and to the in-band sub-band points (dashed blue line) are shown; both fits use at least four flux measurements per source. The fitted spectral indices agree well and the model curves pass through the external measurements. For the brighter sources, the external catalogues are broadly consistent with the MeerKLASS fits. For the faintest $\sim 0.8$\,mJy source in the right-hand panel, however, several external measurements lie close to their survey limits and have visibly larger fractional uncertainties, while the MeerKLASS L-band and sub-band points still have small error bars.  This demonstrates that MeerKLASS delivers higher SNR flux densities at mJy levels than the comparison surveys. The combination of this depth with multiple in-band SED samples allows us to derive spectral indices directly from MeerKLASS data, and to reduce the influence of cross-survey calibration offsets and long-term variability on our spectral characterisation.

\begin{figure*}
    \centering
    \includegraphics[width=1\linewidth]{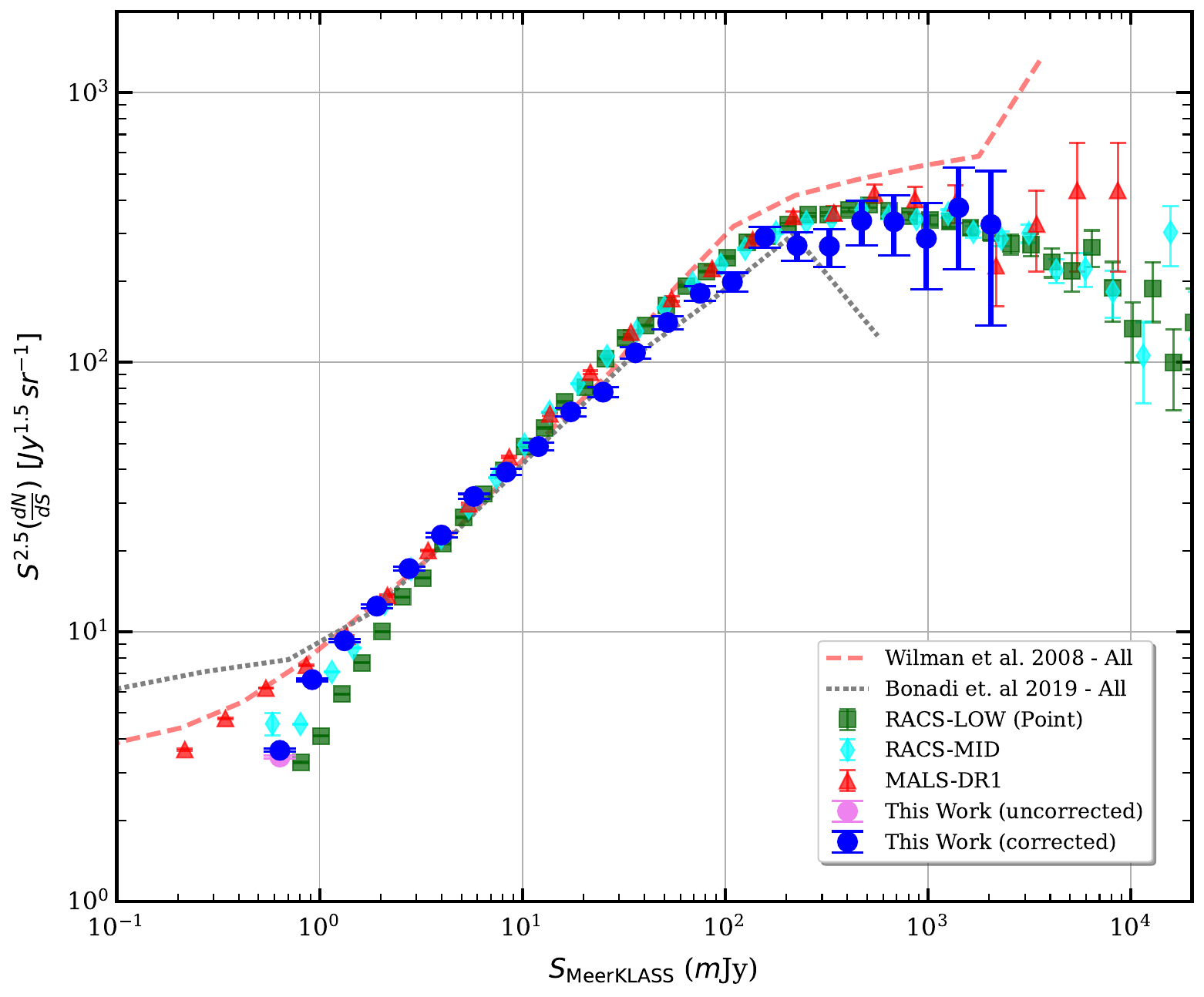}
    \vskip-8pt
    \caption{Euclidean-normalised differential source counts, $S^{2.5}dN/dS$, for the MeerKLASS L-band DR1 catalogue (blue circles: completeness-corrected) shown in comparison with RACS-low, -mid and MALS-DR1 counts, scaled to the reference frequency using $\alpha = -0.9$, $\alpha = -0.7$ and $\alpha = -0.75$ respectively.}
    \label{fig:DNDS}
\end{figure*}

\subsection{Source counts}
\label{sec:sourcecounts}

\subsubsection{Completeness}

To quantify the efficiency of our detection process and establish the survey completeness (the probability of detecting a source as a function of its flux density), we performed a comprehensive set of injection-recovery simulations across all 67 tiles of the L-band footprint. The measured completeness curves are critical for correcting differential source counts especially for faint populations. \par 

For each image tile, we simulated the presence of artificial point sources by drawing their flux densities from a realistic power-law distribution ($\mathrm{d}N/\mathrm{d}S \propto S^{-1.6}$) spanning the range 0.1 to 100\,mJy. These sources are modeled as elliptical Gaussians matching the specific synthesized beam parameters of the corresponding tile and were injected into the $\texttt{PyBDSF}$ residual maps (in units of $\text{Jy}/\text{beam}$). \par

The analysis utilized 50 independent realizations for each tile, with 1000 sources injected per realization. The images were then re-processed using the identical $\texttt{PyBDSF}$ configuration employed for the original catalogue generation. A source was considered successfully recovered if a detection was found within 3\,arcsec of the injected position. The completeness in each flux bin was subsequently defined as the fraction of injected sources that were recovered. We presents the results of this injection-recovery analysis in \autoref{fig:completness}. The right panel displays the mean completeness curves for three cases, highlighting the impact of integration depth and local noise properties:

\begin{enumerate}
    \item Full DR1 Survey (Blue): The survey-wide average reaches 50\% completeness at approximately 0.25\,mJy and approaches 90\% completeness at 0.55\,mJy.
    \item Central Tiles (Red): This subset of tiles, representing a high-sensitivity region of the mosaic, shows improved performance, achieving 90\% completeness near 0.4\,mJy.
    \item Deepest Tile (Tile 18, Green): This single, clean tile (which is free of bright sources), yields the highest recovery rate, reaching 50\% completeness at approximately 0.2\,mJy and 90\% completeness around 0.25\,mJy.
\end{enumerate}

The left panel of \autoref{fig:completness} illustrates the tile-to-tile variation in completeness (thin grey lines), particularly evident at low flux densities. The solid blue line shows the median completeness curve across all tiles, with the bars indicating the $1\sigma$ spread across the realizations. These differences underscore the influence of local RMS noise on detection efficiency, confirming that areas with enhanced integration time provide substantially higher sensitivity. \par

\renewcommand{\arraystretch}{1.5}
\begin{table*}
    \centering
    \caption{Tabulated source counts for the MeerKLASS L-band DR1}

    \begin{tabular}{ccccc}
    \hline
   Flux $S$  & Flux $S$, centre  & Count & $S^{2.5} dN/dS$ & Corrected $S^{2.5} dN/dS$\\
      (mJy) & (mJy)  &  & ($Jy^{1.5}sr^{-1}$) & ($Jy^{1.5}sr^{-1}$) \\
        \hline
	 0.521 - 0.752 & 0.637 & 5584 & $3.433 \pm 0.045$ & $3.635 \pm 0.048$ \\
	 0.752 - 1.086 & 0.919 & 6318 & $6.651 \pm 0.083$ & $6.651 \pm 0.083$ \\
	 1.086 - 1.567 & 1.327 & 5115 & $9.269 \pm 0.129$ & $9.269 \pm 0.129$ \\
	 1.567 - 2.262 & 1.915 & 3984 & $12.460 \pm 0.197$ & $12.460 \pm 0.197$ \\
	 2.262 - 3.265 & 2.764 & 3176 & $17.174 \pm 0.304$ & $17.174 \pm 0.304$ \\
	 3.265 - 4.713 & 3.989 & 2442 & $22.861 \pm 0.462$ & $22.861 \pm 0.462$ \\
	 4.713 - 6.802 & 5.757 & 1959 & $31.775 \pm 0.718$ & $31.775 \pm 0.718$ \\
	 6.802 - 9.817 & 8.309 & 1392 & $39.138 \pm 1.049$ & $39.138 \pm 1.049$ \\
	 9.817 - 14.168 & 11.993 & 999 & $48.697 \pm 1.540$ & $48.697 \pm 1.540$ \\
	 14.168 - 20.449 & 17.309 & 775 & $65.498 \pm 2.353$ & $65.498 \pm 2.353$ \\
	 20.449 - 29.514 & 24.982 & 529 & $77.515 \pm 3.370$ & $77.515 \pm 3.370$ \\
	 29.514 - 42.598 & 36.056 & 427 & $108.485 \pm 5.250$ & $108.485 \pm 5.250$ \\
	 42.598 - 61.481 & 52.039 & 319 & $140.524 \pm 7.868$ & $140.524 \pm 7.868$ \\
	 61.481 - 88.735 & 75.108 & 236 & $180.258 \pm 11.734$ & $180.258 \pm 11.734$ \\
	 88.735 - 128.070 & 108.402 & 150 & $198.655 \pm 16.220$ & $198.655 \pm 16.220$ \\
	 128.070 - 184.843 & 156.456 & 127 & $291.636 \pm 25.878$ & $291.636 \pm 25.878$ \\
	 184.843 - 266.782 & 225.812 & 68 & $270.754 \pm 32.834$ & $270.754 \pm 32.834$ \\
	 266.782 - 385.045 & 325.913 & 39 & $269.254 \pm 43.115$ & $269.254 \pm 43.115$ \\
	 385.045 - 555.732 & 470.388 & 28 & $335.186 \pm 63.344$ & $335.186 \pm 63.344$ \\
	 555.732 - 802.084 & 678.908 & 16 & $332.108 \pm 83.027$ & $332.108 \pm 83.027$ \\
	 802.084 - 1157.642 & 979.863 & 8 & $287.926 \pm 101.797$ & $287.926 \pm 101.797$ \\
	 1157.642 - 1670.817 & 1414.229 & 6 & $374.433 \pm 152.862$ & $374.433 \pm 152.862$ \\
	 1670.817 - 2411.477 & 2041.147 & 3 & $324.620 \pm 187.420$ & $324.620 \pm 187.420$ \\
     \hline
    \end{tabular}
    \label{tab:dnds}
\end{table*}
\renewcommand{\arraystretch}{1.0}

\subsubsection{Differential source counts}
\label{sec:dn_ds}

The differential source count, $dN/dS$, is a statistical tool for extragalactic radio astronomy, quantifying the number of sources per unit solid angle per unit flux density. It offers crucial insights into the cosmic evolution and demographics of radio source populations, such as AGN and SFGs. By convention, we present the Euclidean-normalized form, $S^{2.5}dN/dS$, where $S$ is the integrated flux density. 

We derive the L-band source counts using the final merged catalogue, focusing on sources with an integrated flux density greater than 0.54\,mJy and a SNR greater than 9 in each of the 67 survey tiles. Source are grouped into logarithmic bins of flux density to reduce the impact of sampling noise.

A key step in this process is correcting for catalogue incompleteness, which is pronounced at flux densities near the survey detection threshold. We apply this correction to the raw counts in each bin by dividing them by the corresponding completeness fraction derived from our detailed injection-recovery simulations. The total uncertainty in each bin was calculated by adding in quadrature the Poisson uncertainties on the detected source counts with the statistical errors derived from the completeness simulations.

The final Euclidean-normalized differential source counts, referenced at 1284\,MHz, are presented in \autoref{fig:DNDS}. The plot shows the raw counts (purple circles) and the final completeness-corrected counts (blue circles). The largest effect of the correction is visible in the faintest point in the flux distribution, where the raw counts drop steeply due to incompleteness.

For validation, we compare our measurements (blue circles) with existing results from the RACS-low and -mid and MALS-DR1 catalogues. To ensure a fair comparison, all literature counts are scaled to the MeerKLASS L-band reference frequency (1284\,MHz) assuming a mean spectral index of $\alpha=$\,-0.9 for RACS-low, $\alpha=$\,-0.7 for RACS-mid and $\alpha=$\,-0.75 for MALS-DR1. The MeerKLASS counts are in good agreement with the broad structure defined by the external surveys, successfully bridging the data gap between the shallower RACS-mid data and the deeper end of the population. The distribution accurately traces the cosmic trend: a steep rise from faint fluxes to a turnover (or ``shoulder'') that marks the transition into the Euclidean regime, followed by a decline at the brightest fluxes due to the rarity of bright luminous sources.

The ability of MeerKLASS to reliably trace the source counts across this wide flux range, reaching down to the flux density regime where the contribution from SFGs and faint AGN becomes dominant, strongly validates the survey's calibration, merging process, and completeness corrections.

\section{Public data release}
\label{sec:public_release}

With this manuscript, we are releasing a comprehensive set of data products to support further scientific investigations utilizing the MeerKLASS L-band survey. This DR1 package consists of images and residual maps for each survey tile (see \autoref{fig:alltiles} together with merged catalogues. All data are provided in \texttt{FITS} format with standard $\texttt{WCS}$ (ICRS, $\text{J}2000$); fluxes are in $\text{Jy}\,\text{beam}^{-1}$, and beam information is included in the headers ($\texttt{BMAJ}/\texttt{BMIN}/\texttt{BPA}$).

\begin{itemize}
    \item For the 67 imaged tiles defined in \autoref{fig:alltiles}, we release the Stokes I continuum image centered at 1284\,MHz with a 1.5$\arcsec$ pixel size, and the accompanying residual image to assess the quality of the sky model and calibration.
    
    \item Source Catalogues: We release two \texttt{FITS} binary tables created using the $\texttt{PyBDSF}$ pipeline:
    
    \begin{enumerate}
        \item Source Catalogue ($\texttt{SRL}$): This table contains the final set of radio sources detected after removal of duplicates in the overlapping tile regions.  There are 34,874 radio sources.
        \item Gaussian-Component Catalogue ($\texttt{GAUL}$): This table lists the individual Gaussian components fitted to model the emission, comprising 40,141 components.
    \end{enumerate}
    
    The complete list of columns for both tables is detailed in Appendix ~\ref{app:column_description}. The $\texttt{Source\_id}$ column should be used as the definitive link between the $\texttt{GAUL}$ and $\texttt{SRL}$ catalogues, while the $\texttt{Tile\_ID}$ provides the necessary information to map to the images for a tile-specific analyses.
\end{itemize}

\section{Conclusions}
\label{sec:Conclusions}
We present the first public data release (DR1) of the MeerKLASS continuum L-band survey, utilizing MeerKAT scanning mode observations that have been processed using an On-The-Fly (OTF) interferometric imaging pipeline. Using approximately 13.5 hours of early-science observations, we have demonstrated the promise of efficient scanning-- needed or single-dish intensity mapping-- to produce commensal interferometric imaging over large areas of the sky. We design and validate a dedicated processing pipeline \citep{Chatterjee_2025_OTF}, specifically tailored to the unique challenges of OTF imaging, yielding high-fidelity continuum images and a comprehensive source catalogue for an area of approximately $\sim$268\,deg$^2$ that largely overlaps with the KiDS-DR5 footprint. We quantify the survey imaging performance through rigorous characterization of image fidelity, angular resolution, flux density accuracy, astrometric precision, and completeness. Extensive injection–recovery simulations are employed to accurately measure the completeness as a function of flux density per tile, enabling robust corrections for our final source counts.

Key Findings and Data Products are as follows:

\begin{enumerate}

    \item We produce and release deep continuum images at a central frequency of 1284\,MHz. The survey, covering $\sim268\,\deg^2$ of sky, reaches a median RMS sensitivity of $\sim 33\,\umu\text{Jy}\,\text{beam}^{-1}$ with a median angular resolution defined by a restoring beam of $\sim25.5\arcsec \times 7.8\arcsec$. The overall image quality is as expected for traditional pointed interferometric observations, with noise properties closely tracking the spatial variation in integration time across the OTF scan pattern (equivalently, $\sigma \propto N_{\rm hit}^{-1/2}$ for repeated passes). 

    \item We extract, validate and release a source catalogue containing 34,874 unique radio sources. Cross-matching with external radio surveys (RACS-low, -mid, -high, TGSS, NVSS, and SUMSS) is carried out to measure systematic positional errors that are sub-arcsecond, well below the 1.5$\arcsec$ pixel scale, and to measure the flux-density accuracy, which we show to be robust, demonstrating systematic shifts in flux scale relative to the comparison surveys that are statistically significant but smaller than 5\%.
    
    \item Using MeerKLASS L-band fluxes together with TGSS, SUMSS, RACS-low, -mid, -high and NVSS, as well as the seven MeerKLASS sub-bands, we fit single power–law SEDs for sources with at least four independent flux-density measurements and formal spectral-index uncertainties $\sigma_\alpha \le 0.2$. The resulting spectral-index distributions from MeerKLASS+Others (2,230 sources), MeerKLASS+RACS(all) (1,865 sources) and the in-band sample (6,237 sources) all peak near $\alpha \simeq -0.8$, consistent with a synchrotron-dominated extragalactic population. The in-band indices ($\alpha_{\rm in\text{-}band}$) agree well with the multi-survey fits and are based entirely on MeerKLASS data, demonstrating that the depth and sub-band sampling of the survey provide competitive spectral-index measurements at the mJy level while reducing sensitivity to cross-survey calibration offsets and long-term variability.

    \item We compute the differential source counts for sources above $0.52\,\text{mJy}$, finding excellent agreement with previous surveys. This result validates the entire processing workflow, from calibration through to completeness correction, demonstrating that this DR1 dataset yields statistically robust measurements of the source population.
    
\end{enumerate}

This DR1 dataset represents only a fraction of the total 10,000\,deg$^2$ planned MeerKLASS coverage, which will proceed in the UHF-band.  The $\sim$800\,deg$^2$ UHF component of MeerKLASS DR1 is reported in a companion paper \citep[see][]{Paul_2025_OTF}. The success of this early L-band and UHF-band survey provides crucial confirmation of the OTF interferometric imaging technique and the data processing chain necessary for future large-scale radio surveys with MeerKAT and the SKA. The L-band source catalogues and images reported here are a valuable public resource, enabling science across galaxy evolution, spectral index mapping, and cross-identification with optical surveys like KiDS and the upcoming DESI-DR11.  

In the upcoming MeerKLASS DR2, scheduled for December 2026, we will release L-band imaging and source catalogues over the same region, but including 5$\times$ more scan observing blocks.  DR2 will also include additional UHF imaging.  A new OTF-optimized scanning mode at MeerKAT has been developed with our support and introduced for use in obtain MeerKLASS and other scan surveys starting now.  This will remove the time-smearing anisotropy in the synthesized beam that is apparent in our DR1 data.

The MeerKLASS survey data acquisition strategy, characterized by constant-elevation scans of rising and setting fields, enables both single-dish intensity mapping science and commensal OTF interferometric imaging.  It serves as a vital proving ground for scalable data processing and analysis workflows, that are required for MeerKLASS and other next-generation wide-area radio surveys. The combined challenges addressed in this work, of managing a massive data volume from continuous scanning, maintaining high image fidelity, and implementing techniques like visibility-domain mosaicking, are directly relevant to the high survey speed operations planned for SKA-Mid. The MeerKLASS methodology helps demonstrate how SKA-Mid can achieve its ambitious goals for large-area surveys that unify both cosmological and astrophysical objectives.

The MeerKLASS program is structured to deliver several increasingly deeper and wider data releases:

\begin{itemize}
    \item MeerKLASS UHF Band: The full MeerKLASS program aims to expand coverage to $10,000\,\text{deg}^2$ in the UHF band. The first data release for the UHF band (DR1) is also described in a companion paper \citep[see][]{Paul_2025_OTF}, covering an early-science footprint.
    
    \item Future L-band Data (DR2): The next L-band release, DR2, will incorporate significantly more data, comprising approximately 41 blocks. This expanded dataset is expected to reach as nearly twice the sensitivity achieved in this current DR1.
    
    \item Pipeline Enhancements: Future releases will benefit from ongoing enhancements to the MeerKAT correlator, which are designed to address the current smearing effects of the OTF tracking system. These improvements aim to significantly enhance the effective angular resolution and point-source sensitivity of the processed images.
\end{itemize}

\section*{Acknowledgements}

SM and JM acknowledge the support provided by the German Federal Ministry of Education and Research (BMBF) through the BMBF D-MeerKAT III award (number 05A23WM2). This funding was allocated via the `Verbundforschung' initiative.  In addition, we acknowledge the hardware support of the DFG supported WAP program at LMU and we offer special thanks to the Rechenbetriebsgruppe within the Faculty of Physics. SC acknowledges financial support from the South African National Research Foundation (Grant No. 84156) and the Inter-University Institute for Data Intensive Astronomy (IDIA). IDIA is a partnership of the University of Cape Town, the University of Pretoria and the University of the Western Cape. IDIA is registered on the Research Organization Registry with ROR ID 01edhwb26, and on Open Funder Registry with funder ID 100031500. SP acknowledges support from the Science and Technology Facilities Council (STFC) through the Consolidated Grant ST/X001229/1 at the Jodrell Bank Centre for Astrophysics, University of Manchester. 

The MeerKAT telescope is operated by the South African Radio Astronomy Observatory, which is a facility of the National Research Foundation, an agency of the Department of Science, Technology and Innovation.

We acknowledge the use of the ilifu cloud computing facility – \url{www.ilifu.ac.za}, a partnership between the University of Cape Town, the University of the Western Cape, Stellenbosch University, Sol Plaatje University and the Cape Peninsula University of Technology. The ilifu facility is supported by contributions from the Inter-University Institute for Data Intensive Astronomy (IDIA – a partnership between the University of Cape Town, the University of Pretoria and the University of the Western Cape), the Computational Biology division at UCT and the Data Intensive Research Initiative of South Africa (DIRISA).

This research made use of \texttt{ASTROPY}, a community-developed core Python package for Astronomy \citep{astropy:2022}, \texttt{NUMPY} \citep{Numpy2020array}, \texttt{SCIPY} \citep{SciPy-NMeth2020}. This research also made use of \texttt{MATPLOTLIB} \citep{matplotlib07} open-source plotting packages for \texttt{PYTHON}.

%%%%%%%%%%%%%%%%%%%%%%%%%%%%%%%%%%%%%%%%%%%%%%%%%%
\section*{Data Availability}

All the radio observation data used in this study are available in the SARAO Online Archive \href{https://archive.sarao.ac.za}{(https://archive.sarao.ac.za)} with proposal ID SCI-20210212-MS-01. \par All data products described in \autoref{sec:public_release} will be made available on the MeerKLASS survey webpage: \url{https://meerklass.org/}.

%%%%%%%%%%%%%%%%%%%% REFERENCES %%%%%%%%%%%%%%%%%%

% The best way to enter references is to use BibTeX:

\bibliographystyle{mnras}
\bibliography{main} % if your bibtex file is called example.bib

% Alternatively you could enter them by hand, like this:
% This method is tedious and prone to error if you have lots of references
%\begin{thebibliography}{99}
%\bibitem[\protect\citeauthoryear{Author}{2012}]{Author2012}
%Author A.~N., 2013, Journal of Improbable Astronomy, 1, 1
%\bibitem[\protect\citeauthoryear{Others}{2013}]{Others2013}
%Others S., 2012, Journal of Interesting Stuff, 17, 198
%\end{thebibliography}

%%%%%%%%%%%%%%%%%%%%%%%%%%%%%%%%%%%%%%%%%%%%%%%%%%

%%%%%%%%%%%%%%%%% APPENDICES %%%%%%%%%%%%%%%%%%%%%

\appendix

\section{CATALOGUE STRUCTURE}
\label{app:column_description}

This data release contains two \texttt{FITS} binary tables: (i) the \emph{source} catalogue (\texttt{MeerKLASS\_Lband\_DR1\_SRL.fits}; hereafter SRL), and (ii) the \emph{Gaussian‐component} catalogue (\texttt{MeerKLASS\_Lband\_DR1\_GAUL.fits}; hereafter GAUL). The imaging was organised into $2.15\degr\,\times\,2.15\degr$ tiles with $\simeq0.075\degr$ overlaps, each identified by the column \texttt{Tile\_ID}. 

\subsection*{The SRL (source) catalogue}
The SRL table lists one row per \emph{source} after cross‐tile de‐duplication. For the Source catalogue, we define the following columns:

\begin{itemize}
  \item \textbf{Source Name} (\textit{string}) - The name of the source given in the IAU convention \texttt{MeerKLASS\_Lband\_DR1\,JHHMMSS.S$\pm$DDMMSS.S}, with the prefix \texttt{MeerKLASS\_Lband\_DR1}\footnote{The DR1 has been added as we named this Data Release 1}. Derived from the best astrometric position in ICRS, with a precision of 0.1\,s in RA and 0.1$\arcsec$ in Dec. 

  \item \textbf{N\_Gaus} \,(\textit{int}) - number of Gaussian components associated with the source.
  
  \item \textbf{Source\_id} \,(\textit{string}) -- globally unique source identifier, constructed as \texttt{Tile\_\{Tile\_ID\}\_\{local Source\_id\}} (e.g. \texttt{Tile\_18\_123}). This serves as the primary key used to join GAUL $\rightarrow$ SRL.
  
  \item \textbf{PyBDSF block} -- all native PyBDSF source columns from \textbf{RA} through \textbf{S\_Code} (inclusive), in PyBDSF’s original order. Typical fields include: \texttt{RA, E\_RA, DEC, E\_DEC} in degree; \texttt{Total\_flux, E\_Total\_flux} in Jy; \texttt{Peak\_flux, E\_Peak\_flux} in Jy\,beam$^{-1}$, deconvolved sizes/PA, local noise (\texttt{Isl\_rms}), and the PyBDSF source code \texttt{S\_Code} (e.g. single/multiple).
  
  \item \textbf{E\_Total\_flux\_combined} -- The total flux-density error obtained by adding in quadrature the PyBDSF fitting error and the flux-density terms from Section 6.2 of \citet{Chatterjee_2025_OTF}. 
  
  \item \textbf{Tile\_ID} \,(\textit{int}) -- integer tile identifier of the retained detection after de‐duplication.
  
  \item \textbf{Tile\_BMAJ}, \textbf{Tile\_BMIN} \,(arcsec), \textbf{Tile\_BPA} \,(deg) -- restoring‐beam FWHM major/minor axes and position angle for the corresponding tile, propagated from the tile image headers (\texttt{FITS} keywords \texttt{BMAJ/BMIN/BPA}).

\end{itemize}

\subsection*{The GAUL (Gaussian components) catalogue}
The GAUL table lists individual 2D Gaussian components fitted by PyBDSF and retained for the SRL sources (components are taken from the same tile as the winning SRL entry; no additional de‐duplication is applied at the component level). Key columns are:

\begin{itemize}\itemsep0.2em
  \item \textbf{Source Name} (\textit{string}) -- The name of the source given in the IAU convention \texttt{MeerKLASS\_Lband\_DR1\,JHHMMSS.S$\pm$DDMMSS.S}, with the prefix \texttt{MeerKLASS\_Lband\_DR1}. Derived from the best astrometric position in ICRS, with a precision of 0.1\,s in RA and 0.1$\arcsec$ in Dec. 
  
  \item \textbf{Gaus\_id} \,(\textit{string}) -- globally unique component identifier \texttt{Tile\_\{Tile\_ID\}\_\{local Gaus\_id\}}.
  
  \item \textbf{Source\_id} \,(\textit{string}) -- parent SRL key, same format as in SRL (above). This is the foreign key for GAUL$\rightarrow$SRL joins.
  
  \item \textbf{Tile\_ID} \,(\textit{int}) -- tile identifier.
\end{itemize}

The GAUL table otherwise contains the standard PyBDSF component parameters: component centre (RA, Dec), peak and integrated flux densities, deconvolved major/minor axes and position angle, and shape parameters. The housekeeping fields \texttt{Isl\_id} and \texttt{Wave\_id} have been removed in the release version for clarity. Units follow PyBDSF conventions (positions in deg, peak in $\rm Jy\,beam^{-1}$, integrated in Jy, sizes in arcsec, angles in deg).

More information on how the parameters in the source (*srl.fits) and Gaussian component (*gaul.fits) catalogues {are} produced by \texttt{PyBDSF} can be found through the \texttt{PyBDSF} documentation\footnote{\url{https://www.astron.nl/citt/pybdsf/write_catalogue.html\#definition-of-output-columns}.}.

\section{Effect of OTF time averaging on flux-density errors}

\label{app:noise_psf}
%%%%%%%%%%%%%%%%%%%%%%%%%%%%%%%%%%%%%%%%%%%%%%%%%%

Because the delay centre is held fixed in azimuth and elevation during the OTF scans, time averaging introduces a non-standard relation between the map noise level and the flux-density uncertainty of a point source. Over each 2\,s integration the complex visibilities that are averaged do not, in general, have the same phase, so some coherence is lost. The amount of coherence loss depends on the fringe rate, which is proportional to the $u$–coordinate. In the $u-v$ plane, the visibility amplitude of a point source is therefore attenuated by a sinc–like function of $u$. In the image plane this is equivalent to convolving the synthesized beam with a top-hat kernel in right ascension, producing the observed smearing of compact sources.

The visibility noise behaves differently. The noise contributions averaged within each 2\,s sample have independent random phases, so any phase rotation during the integration does not change the expected noise amplitude. In the $u-v$ plane the noise amplitude is therefore unaffected by the smearing, and in the image plane the noise point–spread function is set solely by the weighted $u-v$ sampling; it is not convolved with the top-hat in right ascension. As a result, the solid angle of the point–source response is larger than the solid angle of the noise point–spread function on the map. A flux measurement for a point source thus contains several independent noise contributions (`noise pixels'), and the corresponding flux-density error exceeds the local map noise. The increase in the error is equal to the square root of the number of independent noise contributions, i.e. the square root of the ratio of the source point–spread–function solid angle to that of the noise. We account for this effect when computing the catalogue flux-density uncertainties, so the reported flux errors are systematically larger than the local map RMS.

Following the detailed calibration of MeerKLASS OTF data by \citet{Chatterjee_2025_OTF}, we estimate that the ratio between the source and noise solid angles in our L-band DR1 images is $\simeq 1.5$. In practice, we include this by scaling the local island RMS when defining the signal-to-noise ratio and when propagating flux-density errors. Throughout this work we therefore adopt
\[
{\rm SNR} = \frac{S_{\rm peak}}{1.5\,\sigma_{\rm isl}},
\]
where $S_{\rm peak}$ is the peak flux density and $\sigma_{\rm isl}$ is the local island RMS from \texttt{PyBDSF} (see Section~\ref{sec:unresolved}). The empirical relation between SNR and fractional flux scatter, and the corresponding systematic term added in quadrature to the fitting error, follow the flux–error analysis \citep[][Section 6.4]{Chatterjee_2025_OTF} .

% Don't change these lines
\bsp	% typesetting comment
\label{lastpage}
\end{document}